\documentclass[11pt,a4paper]{article}
\pdfoutput=1
\usepackage{jcappub, slashed}

\usepackage{bm,graphicx,amssymb}
\usepackage{color}
\usepackage{tikz}

\definecolor{armygreen}{rgb}{0.29, 0.33, 0.13}

\begin{document}

\title{Probing heavy dark matter decays with multi-messenger astrophysical data}
\author[a]{Koji~Ishiwata,}
\author[b,c]{Oscar~Macias,}
\author[c,b]{Shin'ichiro~Ando}
\author[d]{and Makoto~Arimoto}

\affiliation[a]{Institute for Theoretical Physics, Kanazawa
  University, Kanazawa 920-1192, Japan}

\affiliation[b]{Kavli Institute for the Physics and Mathematics of the
  Universe (WPI), University of Tokyo, Kashiwa, Chiba 277-8583, Japan}

\affiliation[c]{GRAPPA Institute, University of Amsterdam, 1098 XH
  Amsterdam, The Netherlands}

\affiliation[d]{Faculty of Mathematics and Physics, Institute of
  Science and Engineering, Kanazawa University, Kanazawa 920-1192,
  Japan}

\emailAdd{s.ando@uva.nl}
\emailAdd{arimoto@se.kanazawa-u.ac.jp}
\emailAdd{ishiwata@hep.s.kanazawa-u.ac.jp}
\emailAdd{oscar.macias@ipmu.jp}

\abstract{We set conservative constraints on decaying dark matter
  particles with masses spanning a very wide range ($10^4-10^{16}$
  GeV). For this we use multimessenger observations of cosmic-ray (CR)
  protons/antiprotons, electrons/positrons, neutrinos/antineutrinos
  and gamma rays. Focusing on decays into the $\bar{b}b$ channel, we
  simulate the spectra of dark matter yields by using the
  Dokshitzer-Gribov-Lipatov-Altarelli-Parisi equations and the
  \texttt{Pythia} package. We then propagate the CRs of dark matter
  origin till Earth by using the state-of-the-art numerical frameworks
  \texttt{CRPropa}, \texttt{GALPROP} and \texttt{HelMod} for the
  solution of the CR transport equation in the extragalactic, Galactic
  region and the heliosphere, respectively. Conservative limits are
  obtained by requiring that the predicted dark matter spectra at
  Earth be less than the observed CR spectra. Overall, we exclude dark
  matter lifetimes of $10^{28}$~s or shorter for all the masses
  investigated in this work. The most stringent constraints reach
  $10^{30}$~s for very heavy dark matter particles with masses in the
  range $10^{11}$--$10^{14}$~GeV. }

\maketitle
 
\section{Introduction}
\label{sec:intro}    
\setcounter{equation}{0}

The standard model of cosmology is very successful in
explaining the history and evolution of the Universe.
Precise observations of the cosmic microwave background (e.g.,
\cite{Aghanim:2018eyx}), primordial abundances of heavy isotopes
created by Big Bang nucleosynthesis (e.g., \cite{Cyburt:2015mya}), and
other measurements at smaller scales indicate that the energy budget
of the Universe is dominated by dark matter (DM) and dark
energy. However, their fundamental nature has not been identified yet.
In the case of DM, many different attempts have been made to unravel
its particle nature. For example, direct detection experiments are
aiming to detect nuclear recoil events caused by a specific type of DM
candidate called weakly interacting massive
particles~\cite{Aprile:2018dbl}. However, it is difficult to measure a
positive signal with this search method if DM is feebly interacting
with standard model particles or is extremely heavy. It is also
possible that DM has a lifetime that is longer than the age of the
Universe, in which case the DM lifetime cannot be measured at direct
detection experiments or collider searches. These difficulties could
potentially be overcome by detecting yields of DM annihilation or
decay with the use of cosmic ray (CR) detectors. Indirect detection
experiments could thus play a complementary role to other search
strategies in our quest for the discovery of DM particles.

In this paper we search for potential DM signatures in a variety of
archival CR data. We focus on heavy DM candidates whose mass ranges
between $\sim 10^4$ and $10^{16}\,{\rm GeV}$ assuming a finite DM
lifetime.  Such (ultra)heavy DM was proposed in the literature~\cite{Chung:1998zb,Kuzmin:1998uv,Chung:1999ve,Chung:2001cb,Kolb:2007vd,Fedderke:2014ura}. An interesting candidate is decaying gravitino in supergravity model. The CRs from decaying dark matter with TeV scale mass have been studied (see e.g., Refs.\,\cite{Takayama:2000uz,Ibarra:2007wg,Ibarra:2008qg,Ishiwata:2008cu,Covi:2008jy,Ishiwata:2008cv,Ishiwata:2009vx,Ishiwata:2009dk,Buchmuller:2009xv,Ishiwata:2010am} for earlier works), and recently Ref.\,\cite{Dudas:2018npp} have extended the study for heavier gravitino whose mass is around EeV. When the DM mass is much larger than $\sim 1$ TeV, various
particles are produced as the result of fragmentation processes,
including electroweak cascades. This leads to the production of stable
particles such as $p$, $\bar{p}$, $\gamma$, $e^{\pm}$, $\nu$ and $\bar{\nu}$ that in turn diffuse out from
their sources to our detectors. While propagating, CRs undergo several
interactions in the Galactic and extragalactic regions. For example,
Galactic CRs interact with the interstellar gas, ambient photons and
magnetic fields in the interstellar medium.  In addition,
extragalactic CR protons and anti-protons (photons, electrons and
positrons) experience additional photo-hadronic processes
(electromagnetic cascades) by interacting with the background photon
fields, including the cosmic microwave background (CMB) and
extragalactic background light (EBL).  It will be shown that each CR
species from DM has a characteristic spectrum in the energy range of
$10^{-3}$ to $10^{16}\,{\rm GeV}$ that could in principle be detected
in archival CR data. There are some works that have a similar aim to our current study (see {\it e.g.},
Refs.\,\cite{Esmaili:2012us,Murase:2012xs,Murase:2015gea,Ahlers:2013xia, Cohen:2016uyg,Kalashev:2016cre,Aloisio:2015lva,Kachelriess:2018rty,Sui:2018bbh}). However, to the best of our
knowledge, self-consistent simulations of the propagation of all the
stable particles in the energy range of $10^{-3}$ to $10^{16}\,{\rm
  GeV}$ in both the Galactic and extragalactic regions have not been
attempted yet.

\begin{table}[t]
 \begin{center}
  \caption{\small Observations of cosmic-ray particles which are used
    in the analysis. The fourth column shows whether each experiment
    detected the corresponding CRs. Otherwise, the last column shows the
    confidence level (CL) of the upper limits quoted in the references.}
  \begin{tabular}{lccccc}
    \hline \hline
    CRs & Observations & Energy [GeV] & Detected & CL upper limits  \\
    \hline
    Gamma ($\gamma$)  & Fermi-LAT\,\cite{Ackermann:2014usa} & $10^{-1}$\,--\,$10^3$  & \checkmark &   \\
    &CASA-MIA\,\cite{Chantell:1997gs}  & $10^{5}$\,--\,$10^{7}$  & & 90\%  \\
    &KASCADE\,\cite{Apel:2017ocm}  & $10^{5}$\,--\,$10^{7}$  & & 90\%  \\
    &KASCADE-Grande\,\cite{Apel:2017ocm}  & $10^{7}$\,--\,$10^{8}$  & & 90\%  \\
    &PAO\,\cite{Aab:2015bza,Aab:2016agp}  & $10^{9}$\,--\,$10^{10}$  & & 95\%  \\
    &TA\,\cite{Abbasi:2018ywn} & $10^{9}$\,--\,$10^{11}$  & & 95\%  \\
    \hline
    Proton ($p$) &PAO\,\cite{Aab:2017njo}  & $10^{9}$\,--\,$10^{11}$  & \checkmark & 84\%  \\
    \hline
    Anti-proton ($\bar{p}$) &PAO\,\cite{Aab:2017njo}  & $10^{9}$\,--\,$10^{11}$ & \checkmark & 84\%  \\
    &AMS-02\,\cite{Aguilar:2016kjl}  & $10^{-1}$\,--\,$10^{2}$ & \checkmark &  \\
    \hline
    Positron ($e^{+}$) & AMS-02\,\cite{Aguilar:2019owu} & $10^{-1}$\,--\,$10^{3}$ & \checkmark & \\
    \hline
    Neutrino ($\nu$) &IceCube\,\cite{Kopper:2017zzm}  & $10^{5}$\,--\,$10^{8}$ & \checkmark & 90\%  \\
    &IceCube\,\cite{Aartsen:2016ngq}  & $10^{6}$\,--\,$10^{11}$ & & 90\%  \\
    &PAO\,\cite{Aab:2017njo}  & $10^{8}$\,--\,$10^{11}$ & & 90\%  \\
    &ANITA\,\cite{Gorham:2019guw}  & $10^{9}$\,--\,$10^{12}$ & & 90\%  \\
    \hline \hline
  \end{tabular}
  \label{tab:obs}
 \end{center}
\end{table}

Here we simulate the production and propagation of DM decay yields,
including $p$, $\bar{p}$, $\gamma$, $e^{\pm}$,
$\nu$ and $\bar{\nu}$, in the Galactic and extragalactic
regions.  Various types of CRs have been observed in a wide energy
range; MeV--TeV $\gamma$, $\bar{p}$ and $e^{+}$ with
Fermi-LAT~\cite{Ackermann:2014usa} and
AMS-02~\cite{Aguilar:2016kjl,Aguilar:2019owu}, respectively;
in the PeV energy range, photons are observed or
constrained with, {\it e.g.}, KASCADE~\cite{Antoni:2005wq},
KASCADE-Grande~\cite{Apel:2013uni,Apel:2017ocm},
CASA-MIA~\cite{Chantell:1997gs,Glasmacher:1999id},
CASA-BLANCA~\cite{Fowler:2000si}, and DICE~\cite{Swordy:1999um}.
Furthermore, for energies in the EeV range, photon flux upper limits
have been obtained by (PAO)~\cite{Aab:2015bza,Aab:2016agp} and
Telescope Array
(TA)~\cite{Abu-Zayyad:2013qwa,Tsunesada:2017aaq,Abbasi:2018ywn}. Astrophysical $\nu$ have been observed/constrained by
IceCube~\cite{Kopper:2017zzm,Aartsen:2016ngq}, Pierre Auger
Observatory (PAO)~\cite{Aab:2017njo}, and ANITA~\cite{Gorham:2019guw}.
The unprecedented high quality of the publicly available
multi-messenger data described above will allow us to impose robust
constraints on the DM lifetime in a very wide DM mass range. A list
with the CR particles assumed in our analysis is given in
Table~\ref{tab:obs} along with the corresponding references that we
use to extract the data.

This paper is organized as follows. Section~\ref{sec:simulation} presents the computation of the DM decaying spectra for the different CR species and the model frameworks for the solution of the CR transport equation in the extragalactic, Galactic region and Heliosphere, respectively. In Sec.~\ref{sec:results} we show the predicted CR spectra after propagation and the resulting limits on the DM lifetime. Finally, we conclude in Sec.~\ref{sec:conclusions}.

\begin{figure}
 \begin{center}
   \includegraphics[width=12cm]{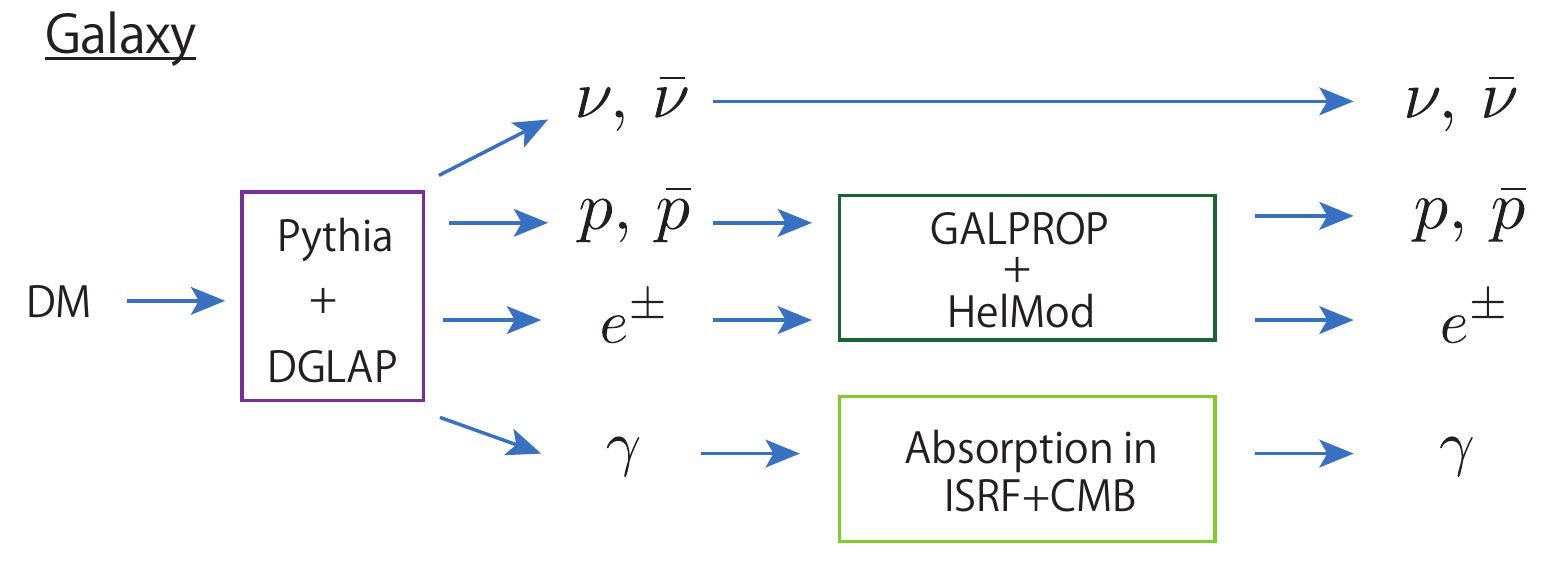}
   \includegraphics[width=12cm]{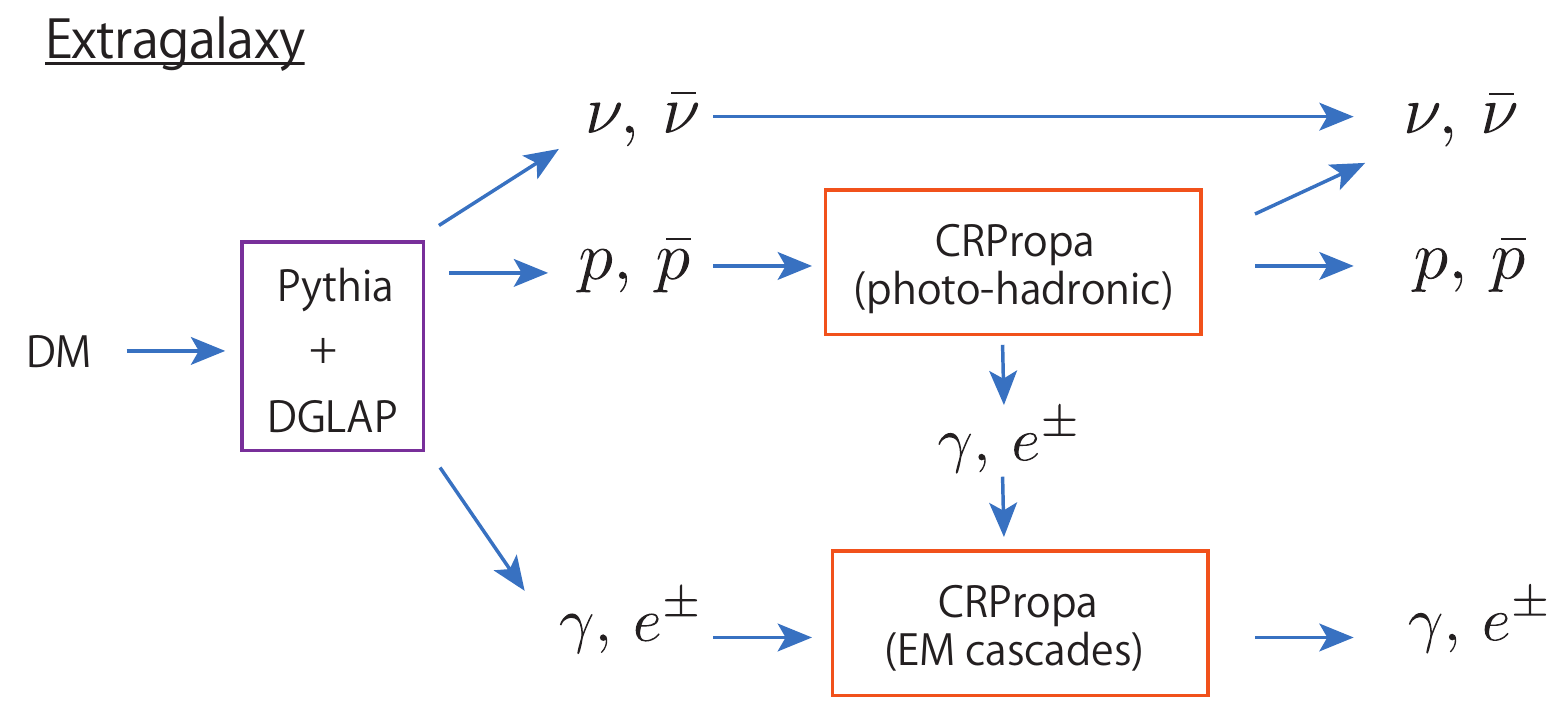}
     \caption{Flowchart of our simulations. Shown are the CR particles
       under consideration in this analysis and the steps carried out
       to propagate those from the DM source till our detectors on
       Earth. The top and bottom panels show that the solution to the
       particle transport equation is done with different methods in
       the Galactic and extragalactic regions. }
  \label{fig:simulation}
 \end{center}
\end{figure}

\section{Cosmic rays from heavy dark matter}
\label{sec:simulation}    
\setcounter{equation}{0}

The predicted CR spectrum from DM decays (at source) is given by the
product of two factors: one that encapsulates the particle physics
properties of the DM candidate and another that gives account of the
abundance and distribution of the DM. This is written as
\begin{equation}\label{eq:spectraatsource}
  \frac{d\Phi_{X}(E_X,\psi)}{dE_{X}} =
  \left( \frac{1}{4\pi \tau_{\rm dm} m_{\rm dm}}
  \frac{dN_{X}}{dE_X}\right)
  \left(\frac{1}{\Delta \Omega}\int_{\Delta\Omega} d\Omega \int_{\rm l.o.s}
  dl \rho_{\rm dm}(r(l,\psi))\right),
\end{equation}
where $m_{\rm dm}$ is the DM mass, $\tau_{\rm dm}$ the DM lifetime, $r(l,\psi)$ is the Galactocentric distance, $l$ and $\psi$ are the distance and direction measured along the line of sight, respectively. $dN_{X}/dE_X$ is the CR spectrum of stable particle $X$ at source, with $X=p$, $\bar{p}$, $e^{\pm}$, $\gamma$, and $\nu$,
$\bar{\nu}$. 

For local DM energy density $\rho_{\rm dm}$, we adopt the spherically
symmetric Navarro-Frenk-White (NFW) profile:
\begin{equation}
\rho(r) = \frac{\rho_s}{(r/r_s)(r/r_s+1)^2},
\end{equation}
where we select $r_s = 11$ kpc, $\rho_{\odot} = 0.43 $ GeV/cm$^3$ and
$R_{\odot} = 8.34 $ kpc for the scale radius and local DM density. We
extract these parameters by inspection of Fig. 6 in
Ref.~\cite{Karukes:2019jxv}.\footnote{We have checked
    that the gamma-ray intensity does not change significantly if
    other halo profile is adopted.  For example, we find a $\sim 10$ \%
    difference if a Burkert profile~\cite{Burkert:1995yz} is used. }

For definiteness, here we consider a scenario in which DM
decays into $b \bar{b}$ final states with a branching ratio of 100\%.
Our simulations are performed in two steps: First, we compute the CR
spectra at source for $p$, $\bar{p}$, $e^\pm$, $\gamma$ and $\nu$,
$\bar{\nu}$ from prompt DM decays. Second, we propagate these
particles in the Galactic and extragalactic medium to derive
observable spectra. A flowchart of our simulations is displayed in
Fig.\,\ref{fig:simulation}.

\subsection{Computation of Cosmic Ray spectra at source}
\label{sec:pytha+DGLAP}

The energy spectra at source of the stable particles resulting from
decaying DM can be computed using the \texttt{Pythia
  8.2}~\cite{Sjostrand:2014zea}. This is the standard method followed
by most studies in the literature. However, this method can be highly
computationally expensive, specially when the DM mass is
larger than $\sim 10$~PeV (in the case of $b \bar{b}$ final state
particles). Due to this technical limitation, in this work we predict
the CR spectra at source using a hybrid approach.  Namely, we use the
Dokshitzer-Gribov-Lipatov-Altarelli-Parisi (DGLAP) equations for the
quantum chromodynamics (QCD) calculations involving DM yields with
$m_{\rm dm}\ge 100$~PeV, otherwise we use the \texttt{Pythia} package. The
procedure to solve the DGLAP equations consists of two parts;
calculation of fragmentation functions (FFs) of hadrons $h=\pi^\pm$,
$\pi^0$, $K^\pm$, $K^0$, $\bar{K}^0$, $n$, $\bar{n}$ and $p$,
$\bar{p}$ (using DGLAP equations); and calculation of the energy
spectra of stable particles resulting from the decay of unstable
hadrons (using \texttt{Pythia}). Similar attempts have been made in earlier
studies using \texttt{HERWIG}~\cite{Bahr:2008pv,Bellm:2015jjp}, the
QCD event generator~\cite{Birkel:1998nx}, \texttt{HERWIG} and FFs in
(supersymmetric) QCD ~\cite{Sarkar:2001se}, Monte Carlo simulation and
FFs in (supersymmetric) QCD~\cite{Berezinsky:2000up,Aloisio:2003xj}
and FFs in the (supersymmetric) standard
model~\cite{Barbot:2002ep,Barbot:2002gt}.

\begin{figure}[t!]
 \begin{center}
   \includegraphics[width=7.5cm]{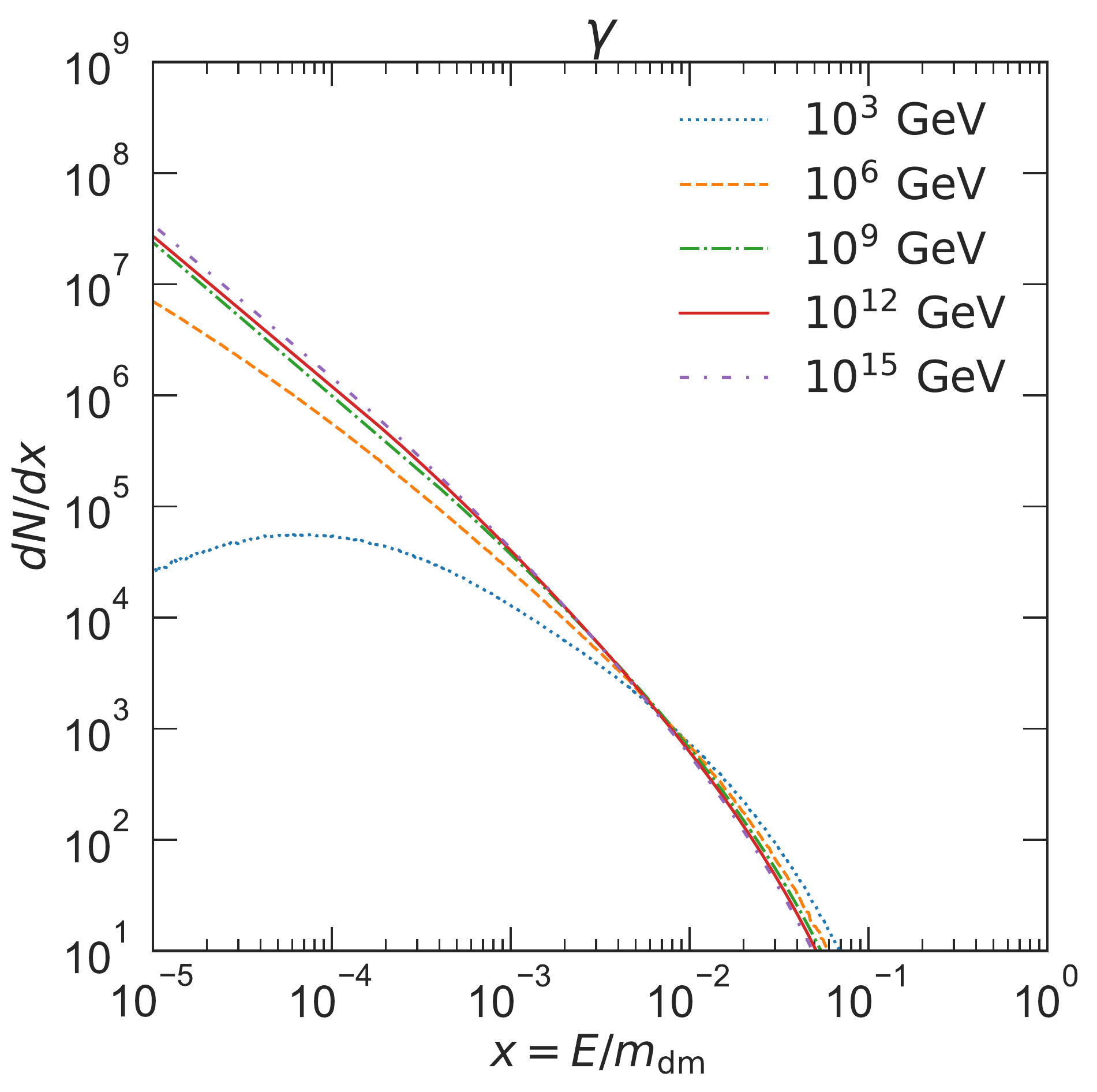}
   \includegraphics[width=7.5cm]{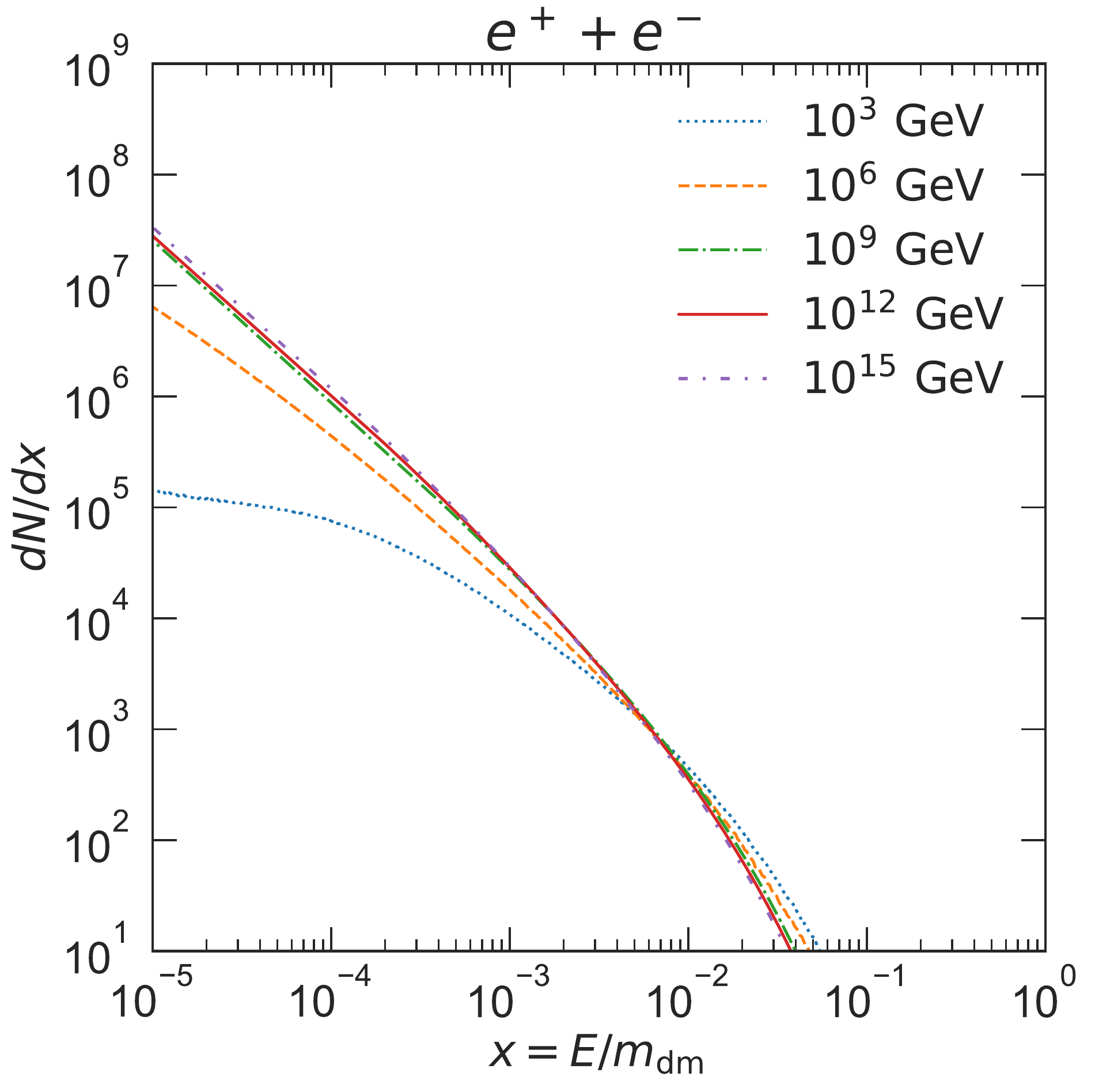}
   \includegraphics[width=7.5cm]{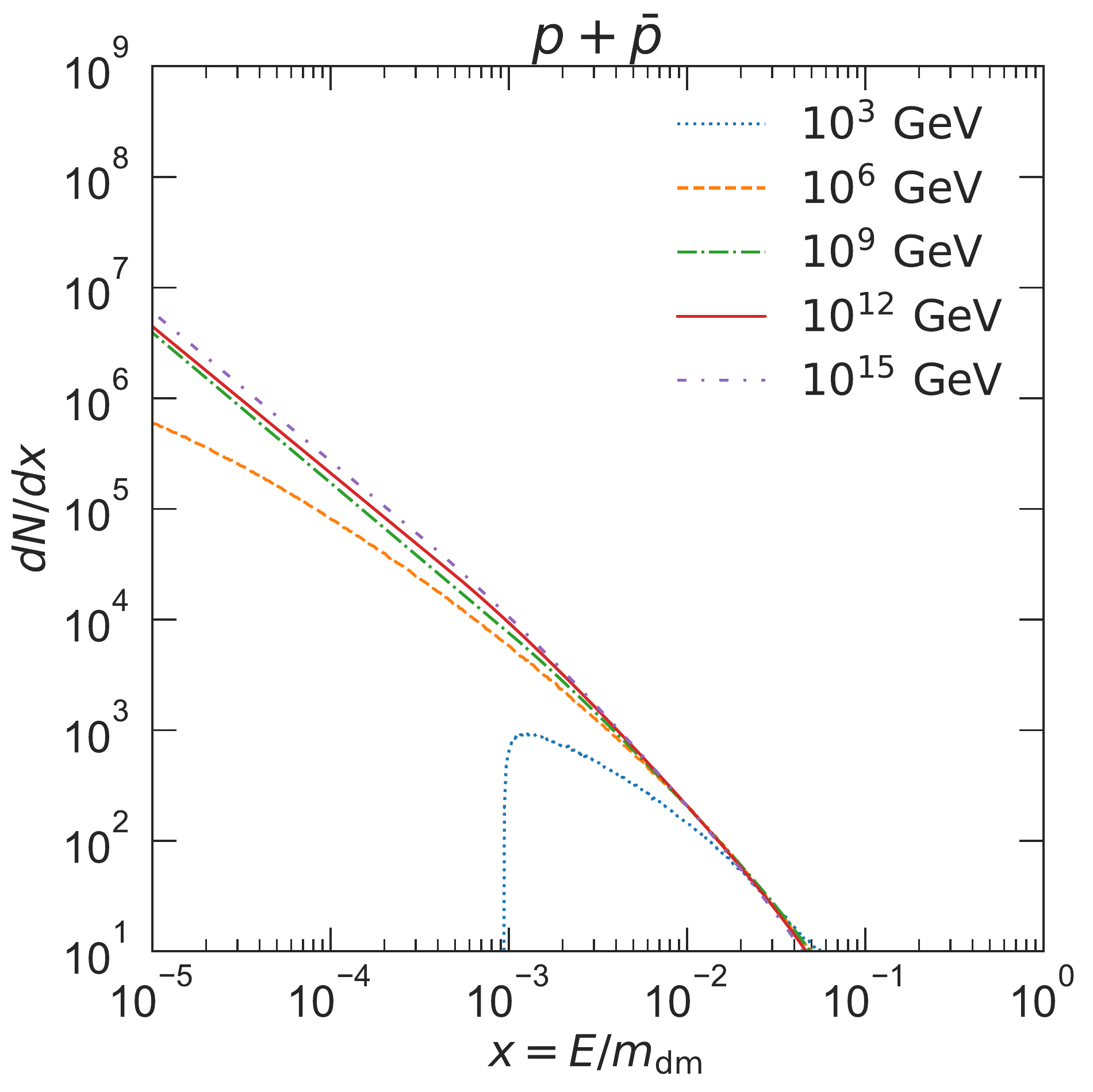}
   \includegraphics[width=7.5cm]{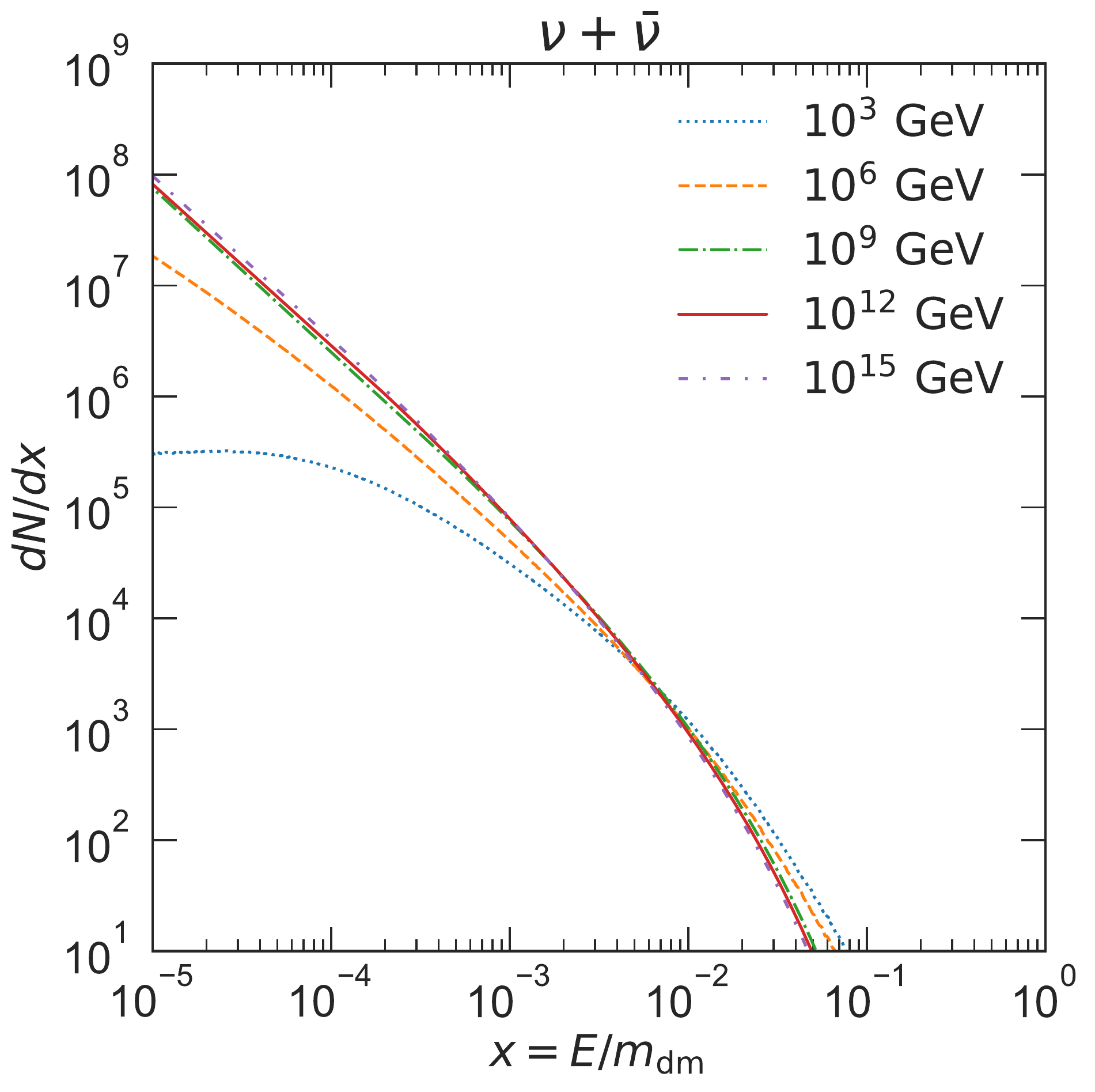}
  \caption{Decaying DM spectra at source $dN/dx$ 
  for $\gamma$, $e^++e^-$,
    $p+\bar{p}$, and $\nu+\bar{\nu}$ as function of $x=E/m_{\rm dm}$. Dark matter is assumed to decay into the $b\bar{b}$ channel. Dark matter masses are
    taken to be $10^3$, $10^6$, $10^9$, $10^{12}$, $10^{15}$\,GeV, respectively.}
  \label{fig:dNdx}
 \end{center}
\end{figure}

The fragmentation functions $D^h_i(z,Q^2)$ of the hadron $h$ for a given 
parton $i$ with energy fraction $z$ are calculated by solving the DGLAP
equations. Currently the next-to-leading order (NLO) results in the
$\overline{\rm MS}$ scheme are
available in~\cite{Kniehl:2000fe,Kretzer:2000yf,Albino:2005me}, and the
uncertainties of parton distribution functions are provided by~\cite{Pumplin:2001ct,Pumplin:2002vw,Martin:2002aw,Martin:2003sk,Bluemlein:2002be,Hirai:2003pm,Leader:2005ci,deFlorian:2005mw,Hirai:2004wq}. In
our study we use the code made available in Refs.~\cite{Hirai:2007cx,Hirai:2011si}.

The energy spectra $f^I_h(x)$ of stable particles ($I=p$, $e$,
$\gamma$, $\nu$) from unstable hadrons $h$ with energy fraction $x$
are calculated with \texttt{Pythia}. Here $f^I_h(x)$ is normalized to
single hadron decay, and both particles and anti-particles are
counted. We have checked that the results agree with the analytical
results given, {\it e.g.}, in Ref.\,\cite{Kelner:2006tc}, for pion
decay products.

Consequently, the energy spectra of stable particles from DM
decays (in the $b\bar{b}$ channel) are given by
\begin{align}
  \frac{dN_I}{dE_I}=
  \frac{2}{m_{\rm dm}}\frac{dN_I}{dz}\,,
 \end{align}
where $z=2E_I/m_{\rm dm}$ and 
\begin{align}
  \frac{dN_I}{dz}=
  2\sum_h \int^1_z \frac{dy}{y} D^h_b(y,m_{\rm dm}^2)f^I_h(z/y)\,.
  \label{eq:dNdz}
\end{align}
The factor of 2 included in the right-hand side of
Eq.\,\eqref{eq:dNdz} results from taking into account contributions
from both $b$ and $\bar{b}$ final states. Figure\,\ref{fig:dNdx} shows
the predicted spectra from DM decaying into the $b
\bar{b}$. For clarity, all panels display the quantity
$dN_I/dx=2dN_I/dz|_{z=2x}$ (where $x=E_I/m_{\rm dm}$). As it can be
seen, the spectra present asymptotic behavior as the $m_{\rm dm}$
increases. Also, the spectral shape and normalizations are different
for each species under consideration. We have checked that the results
obtained with our hybrid method for $m_{\rm dm}\le 10~$TeV agrees well
with those produced by the PPPC4~\cite{Cirelli:2010xx} package with or
without electroweak corrections.\footnote{It has been pointed out that
  electroweak corrections become important for DM masses $\gtrsim
  10$~TeV~\cite{Ciafaloni:2010ti}. This effect is essential in the
  simulation of stable particles. Specially for leptophilic decaying
  DM because $p$, $\bar{p}$, $e^{\pm}$ and $\gamma$
  are produced even when DM decays to, for example, neutrino
  pairs. However, for hadronic decays, the electroweak corrections
  have minor effects on the spectra.} For larger DM mass values, our
results for $\gamma$ can be compared with, {\it e.g.,}
Refs.\,\cite{Cohen:2016uyg,Kalashev:2016cre}. We have found that the
$\gamma$ spectra shown in the published version of Fig.\,S12 in
Ref.\,\cite{Cohen:2016uyg} are quantitatively different from our
results.\footnote{In private communication with the authors of that
  paper the apparent disagreement has been resolved. They have updated
  their Fig.\,S12 on the arXiv with results that now match our own.
} On the other hand, the $\gamma$ spectra shown in Fig.\,1 of
Ref.\,\cite{Kalashev:2016cre} are almost consistent with ours. We
noticed that they are harder in the low $x$ regime. To explore this
further, we have compared our results using \texttt{Pythia}-only versus those
using DGLAP+\texttt{Pythia} for $m_{\rm dm}<100$~PeV and found that the results
obtained with \texttt{Pythia}-only gave softer spectra in the $x\lesssim
10^{-4}$ region. Although it would be interesting to run a more
in-depth investigation of this discrepancy from a viewpoint of Monte
Carlo simulations versus DGLAP evolution, this is beyond the scope of
our current study. In addition, it is expected that the CR particles
at such a low $x$ will have a minor effect on the observable CR fluxes
at Earth after propagation.

\begin{table}[t!]
  \centering
  \caption{\label{tab:galprop_setup} Main GALPROP propagation
    parameter setup considered in this study. Our baseline
    fore/background model corresponds to the best-fit propagation
    parameter setup obtained in
    Ref.~\cite{Boschini:2017fxq}. Other propagation parameters (that
    impact relatively less the results) are taken from Tab.~2 and 3 of
    that reference.}
      \begin{tabular}{ c c c c c c c }\hline \hline
       $D_0$ & $z_h$ & $V_{\rm Alf}$  &$\delta$ & $V_{\rm convec}$& $dV_{\rm convec}/dz$ \\ 
       (10$^{28}$ cm$^2$ s$^{-1}$) & (kpc) &  (km s$^{-1}$)& &(km s$^{-1}$) & (km s$^{-1}$)  \\ \hline 
         4.3 & 4.0 &  28.6 & 0.395 &12.4 &10.2 \\ \hline \hline
    \end{tabular}
\end{table}

\begin{table}[t!]
  \centering
  \caption{\label{tab:helmod_setup} HelMod propagation parameters considered in this study.}
      \begin{tabular}{ c c c  }\hline \hline
       $\rho_i$ & $K_0$ & $g_{\rm low}$   \\ 
        & (AU$^2$ GV$^{-1}$ s$^{-1}$) &   \\ \hline 
         0.065 & $3\times 10^{-5}$ &  0.4  \\ \hline \hline
    \end{tabular}
\end{table}

\subsection{Propagation of Cosmic Rays in the Galaxy}
\label{sec:GALPROP}

In this section we describe the methods used in this work to model the
propagation of CRs in the Galactic region. The propagation of CR
particles in the Galaxy can be studied with various numerical packages
like \texttt{GALPROP}~\cite{Strong:1998fr} or \texttt{DRAGON}~\cite{Evoli:2008dv}.
However, in the very high energy regime, gamma rays of DM
origin can be attenuated via pair production. As pointed out in
Ref.\,\cite{Moskalenko:2005ng}, gamma rays with energies in the range
0.1--100\,PeV tend to be absorbed by the the interstellar radiation
field (ISRF) through interactions of the form $\gamma \gamma \to
e^+e^-$. Since one of the primary science goals of a propagation code
such as \texttt{GALPROP} has been the study of lower energy gamma ray
observations with Fermi-LAT, it does not contain specific routines
designed to output attenuated gamma ray maps. In the very high energy
range, we follow the prescription given in
Ref.\,\cite{Moskalenko:2005ng} which we implement outside the
\texttt{GALPROP} framework, but using the ISRF data that comes with
that package. This is also outlined in the flowchart of
Fig.\,\ref{fig:simulation}.

For CR particles of energies $\lesssim 10^8$\,GeV, our method consists
of using the propagation packages \texttt{GALPROP v54}\footnote{For
  this part of the analysis we use a customized \texttt{GALPROP}
  version explained in Ref.~\cite{Song:2019nrx}.} and \texttt{HelMod
  v4.0} for the solution of the transport equation in the interstellar
medium and the heliosphere, respectively. At its core,
\texttt{GALPROP} consists of a suite of routines that solve the
particle transport equation via numerical methods. Given a certain CR
source distribution, injection spectrum, boundary conditions and
Galactic structure (e.g. interstellar gas, radiation and magnetic
fields), \texttt{GALPROP} makes detailed predictions of relevant
observables for all CR species. The processes accounted for by
\texttt{GALPROP} include pure diffusion, convection (Galactic winds),
diffusive re-acceleration (diffusion in energy space), energy losses
(ionization, Coulomb scattering bremmstrahlung, inverse Compton
scattering and synchrotron radiation), nuclear fragmentation, and
radioactive decay~\cite{Moskalenko:2005ng}. Measurements of CR
isotopes and spectra of primary and secondary CR species made by
Voyager 1, PAMELA, AMS-02, BESS and other balloon experiments allow
the estimate of some of the most important CR propagation
parameters. For example, the ratio of the CR halo size to the
diffusion coefficient can be obtained from measurements of stable
secondary particles such as Boron. The resulting degeneracy between
the CR halo size and the diffusion coefficient can be alleviated with
the observed abundances of radioactive isotopes such as $^{10}_{4}$Be,
$^{26}_{13}$Al, $^{36}_{17}$Cl and
$^{54}_{25}$Mn~\cite{Moskalenko:2005ng}.

Except for Voyager 1 that since 2012 is streaking through space
outside of the heliosphere, all other indirect or direct CR detectors
reside well within its boundaries. While the \texttt{GALPROP}
framework allows for detailed studies of CR propagation through the
Galaxy, it does not contain tools for the solution of the particle
transport equation in the heliosphere. The spectrum of charged CRs
measured at Earth vary with time according to the solar activity. In
particular, solar modulation effects are expected to be important
mainly for CRs of moderate energies ($\lesssim 30$--50
GeV).\footnote{We note that at this energy level the AMS-02 detector
  has made very precise CR observations which we put to use in the
  present work.}  The \texttt{HelMod} package contains dedicated
routines to robustly model the solar modulation on the Galactic CR
spectra. As Galactic CRs enter the heliosphere their trajectories are
affected by solar wind outflows and corresponding magnetic-field
irregularities. \texttt{HelMod} considers both a macroscopic and small
scale heliospheric magnetic field. The former is given by an
Archimedean spiral and the latter by the irregularities originated in
the solar wind. In particular, \texttt{HelMod} uses Monte Carlo
methods to solve the two-dimensional Parker equation for CR transport
through the heliosphere~\cite{Boschini:2017fxq}. For rigidities
greater than 1\,GV, it assumes a parallel component to the magnetic
field of the diffusion tensor given by:
\begin{equation}\label{eq:solardiffusion}
  K_{||}=\frac{\beta}{3} K_0\left[ \frac{P}{1\text{GV}}
    +g_{\rm low}\right] \left(1+\frac{r}{\text{1 AU}}\right),
\end{equation}
where $\beta=v/c$ with $v$ the particle velocity $c$ the speed of
light, $K_0$ is the diffusion parameter, $P=qc/|Z|e$ is the CR
particle rigidity, $r$ is the heliocentric distance from the Sun and,
$g_{\rm low}$ represents the level of solar activity. It also assumes
that the perpendicular diffusion coefficient is proportional to
$K_{||}$, with their ratio denoted by $K_{\perp,i}/K_{||}=\rho_i$ and
$i$ refers to Cartesian coordinate index. 

In order to propagate energetic Galactic CRs to the Earth, we first
use \texttt{GALPROP} to obtain the local interstellar spectra (LIS)
and its output is subsequently fed into \texttt{HelMod} which allows
us to calculate modulated CR spectra for the particular time periods
in which the AMS-02 observations were carried out. In
Ref.~\cite{Boschini:2017fxq} the two packages were combined to
self-consistently model the LIS for protons, helium and anti-protons
assuming different modulation levels and both polarities of the solar
magnetic field. In that work, a propagation parameter scan was carried
out by optimization of a likelihood function constructed using data
taken by AMS-02, BESS, and PAMELA as well as the predicted spectra of
corresponding CR species. Table~\ref{tab:galprop_setup} displays the
best-fit main \texttt{GALPROP} propagation parameters obtained in that
reference which we adopt as our baseline propagation model
setup.\footnote{We use the default setting for the
    magnetic fields. It produces synchrotron emissions that can be
    another signal of DM, {\it
      e.g.},~\cite{Ishiwata:2008qy,Crocker:2010xc,Cirelli:2016mrc}. Ref.\,\cite{Cirelli:2016mrc}
    shows the conservative (progressive) bounds for DM decaying to
    $b\bar{b}$, $\tau_{\rm dm}\gtrsim 10^{24}\,(10^{26})\,{\rm s}$ for
    $m_{\rm dm}\ge 10\,{\rm TeV}$. It will be seen that this
    constraint (even progressive one) is much weaker that the
    constraints obtained from the gamma-ray observations.  } These
are the parameters that were found to produce the largest effect on
the propagated CR spectrum. Namely, the CR halo height $z_h$ (in
Galactocentric coordinates), diffusion coefficient $D_0$ at reference
rigidity $R_D=4.5$ GV, diffusion slope $\delta$, Alfv\'en velocity
$V_{\rm Alf}$, convection velocity $V_{\rm conv}$ and convection
velocity gradient $dV_{\rm conv}/dz$. Other \texttt{GALPROP}
propagation parameters used in our analysis are as given in Tab.~2 and
3 of Ref.~\cite{Boschini:2017fxq}. In turn, the \texttt{HelMod}
propagation parameter setup assumed in our simulations is displayed in
Tab.~\ref{tab:helmod_setup}.

Of particular relevance to this study is the production and
propagation of CR $\bar{p}$ and $e^\pm$. \texttt{GALPROP} classifies
the $\bar{p}$ produced by our Galaxy as ``secondary'' and
``tertiary'' depending on their origin. Namely, ``secondary $\bar{p}$'' are produced through the inelastic interactions given
by $pp$, $pA$, and $AA$ (where $A$ refers to the atomic number of
heavy nuclei) while ``tertiary $\bar{p}$'' result from inelastic
scattering of $\bar{p}$ at propagation. In the case of
$e^{\pm}$, \texttt{GALPROP} also considers primary $e^-$
which are accelerated in CR sources (e.g., supernova remnants) as well as
secondary $e^\pm$ from the collisions of nuclei with the interstellar media. We use the same parameter
setup shown in Tab.~\ref{tab:galprop_setup} for the computation of the
fore/background $\bar{p}$ as well as those of DM origin. Also,
Ref.~\cite{Boschini:2017fxq} did not include $e^{\pm}$ data in their
MCMC scans. In this sense, we do not expect to obtain a suitable
astrophysical background model for $e^{\pm}$ using
Tab.~\ref{tab:galprop_setup}. In light of this we have opted for using
the same propagation setup for $\bar{p}$ as for $e^{\pm}$ when
propagating $e^{\pm}$ of DM origin but only simulated an astrophysical
background model for $\bar{p}$ particles. It will be detailed in a
later section that this is a conservative assumption as in the
$e^{\pm}$ case we will compute DM constraints by imposing that our DM
predicted fluxes do not saturate $e^{\pm}$ measurements.

We note that $p$ and $\bar{p}$ of energies $\gtrsim
10^8$\,GeV propagate just like neutral particles
and thus we could apply the same propagation methods as for photons
and neutrinos. In this case, we can safely neglect the diffusion
effects. As such, we compute the flux of these CRs at Earth by
computing a line-of-sight integral as is done in
Ref.\,\cite{Aloisio:2015lva}.


\subsection{Propagation of Cosmic Rays in the extragalactic region}
\label{sec:crprop}

Decay products from DM undergo cascading processes in the
extragalactic region during the propagation to Earth. We use
\texttt{CRPropa~3.1}~\cite{Batista:2016yrx,Heiter:2017cev} for the
simulation of such processes. Within the \texttt{CRPropa} framework,
\texttt{SOPHIA}~\cite{Mucke:1999yb} and
\texttt{DINT}~\cite{Lee:1996fp} are assumed for the computation of
photo-hadronic processes and electromagnetic cascades,
respectively. We have customized the original code to include CR
particles from decaying DM. \texttt{CRPropa} is specially
suitable to study the propagation of CR nuclei, photons and
electrons/positrons.

In the case of $p$ and $\bar{p}$, two photo-hadronic processes are
relevant (see also `\texttt{CRPropa}/photo-hadronic' in
Fig.\,\ref{fig:simulation}):
\begin{itemize}
\item Photo-pion production: $p+\gamma_{\rm bg}\to p+\pi$,
\item Pair production (Bethe-Heitler): $p+\gamma_{\rm bg}\to p+e^++e^-$.
\end{itemize}
Here $\gamma_{\rm bg}$ refers to the background photons present in the
extragalactic region. For this component we take into account the CMB
and EBL (using the default model in Kneiske
2004~\cite{Kneiske:2003tx}). Through the two processes mentioned
above, $e^\pm$, $\gamma$ and $\nu$, $\bar{\nu}$ are produced as
secondary CRs. The threshold energies for photo-pion production and
pair production are estimated as $6.8\times 10^{10} ({\rm
  meV}/E_{\gamma_{\rm bg}})~{\rm GeV}$ and $4.8\times 10^{8} ({\rm
  meV}/E_{\gamma_{\rm bg}})~{\rm GeV}$,
respectively~\cite{Heiter:2017cev}, with $E_{\rm bg}$ being the energy
of the background photons. For $p$ with energies above $\sim
10^{11}\,{\rm GeV}$, photo-pion productions becomes the dominant
dissipation process with mean energy-loss length of
10~Mpc~\cite{Stanev:2000fb}. This is the main process of the
Greisen–Zatsepin–Kuzmin (GZK)
effect~\cite{Greisen:1966jv,Zatsepin:1966jv}.  Photodisintegration and
elastic scattering processes, on the other hand, are irrelevant for
$p$, and we have checked that nuclear decays produce negligible
effects on the $p$ propagation.

As for $e^\pm$ and $\gamma$ case, four different electromagnetic
cascading effects need to be taken into account (see also
`\texttt{CRPropa}/EM cascades' in Fig.\,\ref{fig:simulation}),
\begin{itemize}
\item Inverse Compton scattering (ICS): $e^\pm+\gamma_{\rm bg}\to
  e^\pm+\gamma_{\rm bg}$,
\item Triplet pair production (TPP): $e^\pm+\gamma_{\rm bg} \to e^\pm + e^+
  + e^-$,
\item Pair production (PP): $\gamma + \gamma_{\rm bg}\to e^+ + e^-$,
\item Double pair production (DPP): $\gamma + \gamma_{\rm bg}\to e^+ +
  e^- +e^+ + e^-$.
\end{itemize}
For the photon background fields, we assume the default setting in
\texttt{DINT}: CMB, EBL (Stecker 2006 model~\cite{Stecker:2005qs}),
and radio background (Protheroe 1996
model~\cite{Protheroe:1996si}).\footnote{Sometimes the EBL and radio
  background are called IRB and URB in \texttt{CRPropa},
  respectively.}  The impact of each process can be seen in Fig.\,5 of
Ref.\,\cite{Heiter:2017cev}. Regarding $e^\pm$-$\gamma_{\rm bg}$
scattering, ICS (TPP) with the CMB is dominant for energy of $e^\pm$
smaller (larger) than $10^8~{\rm GeV}$. As for $\gamma$-$\gamma_{\rm
  bg}$ scattering, DPP is subdominant compared to PP.  The later is
most relevant in the energy range of $10^5\,{\rm GeV} \lesssim E
\lesssim 10^{11}\,{\rm GeV}$ where the main photon background is again
the CMB. It is clear that interactions with the CMB is the most
relevant process in a wide energy range. The
  inter-galactic magnetic fields, on the other hand, have large
  uncertainties. A lower bound is obtained, {\it e.g.},
  Ref.\,\cite{Finke:2015ona}, that is round $10^{-19}\,{\rm G}$. On
  the other hand, it is shown in Ref.\,\cite{Heiter:2017cev} that the
  synchrotron process becomes subdominant when the magnetic fields are
  smaller than 0.1 nG. Therefore, we conservatively ignore the effects
  of the magnetic fields in our evaluation.

Finally, $\nu$ and $\bar{\nu}$ are produced via the photo-hadronic
interactions in addition to the prompt DM decay. Such high-energy
neutrinos may suffer from resonant absorption
processes~\cite{Gondolo:1991rn}. However, we have found that this has
a negligible effect on the neutrino propagation. Therefore, neutrinos
produced via both the photo-hadronic interactions and the prompt dark
matter decay only get redshifted when they reach Earth.

\section{Results}
\label{sec:results}

\begin{figure}[t!]
 \begin{center}
   \includegraphics[width=7.5cm]{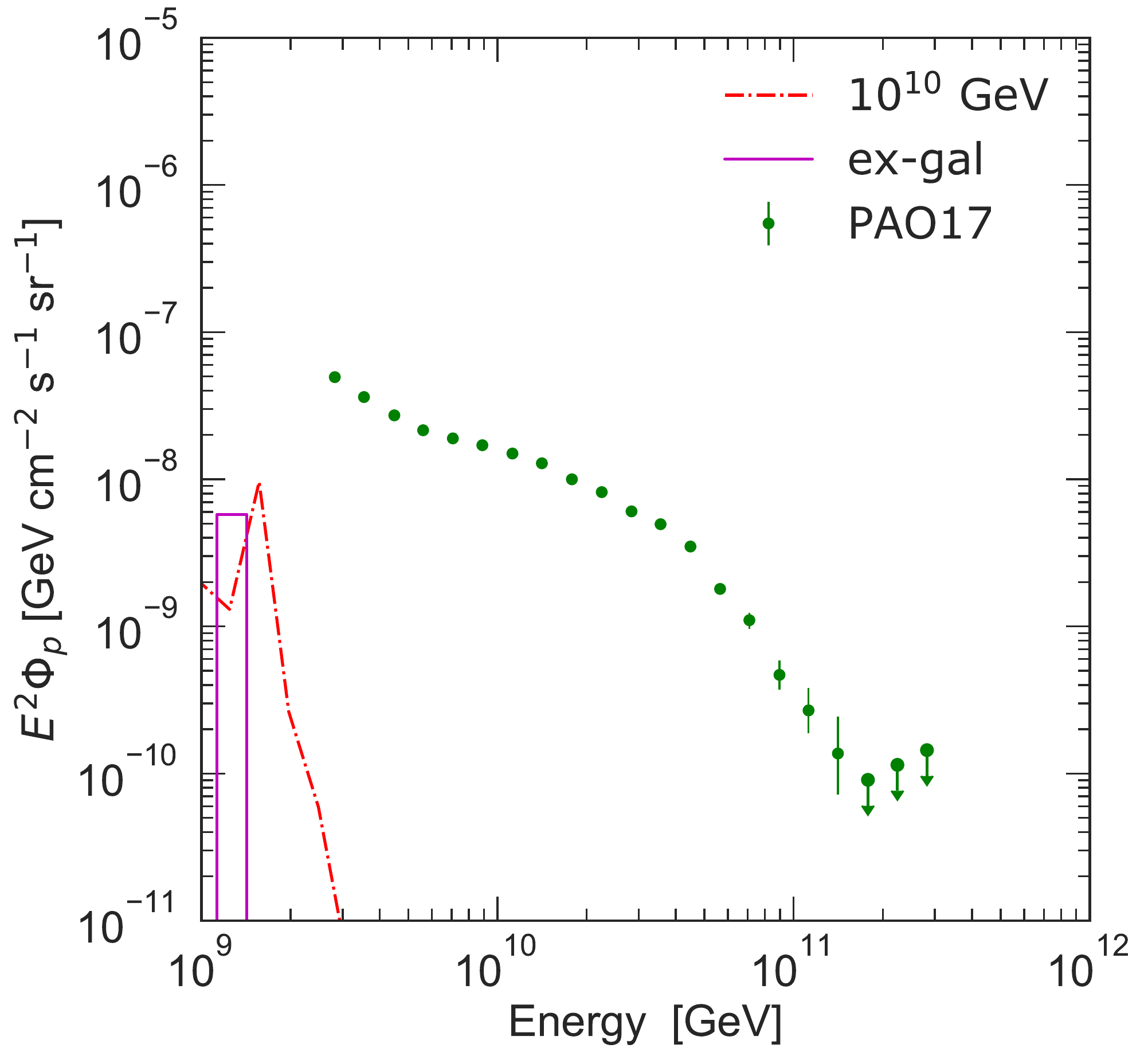}
   \includegraphics[width=7.5cm]{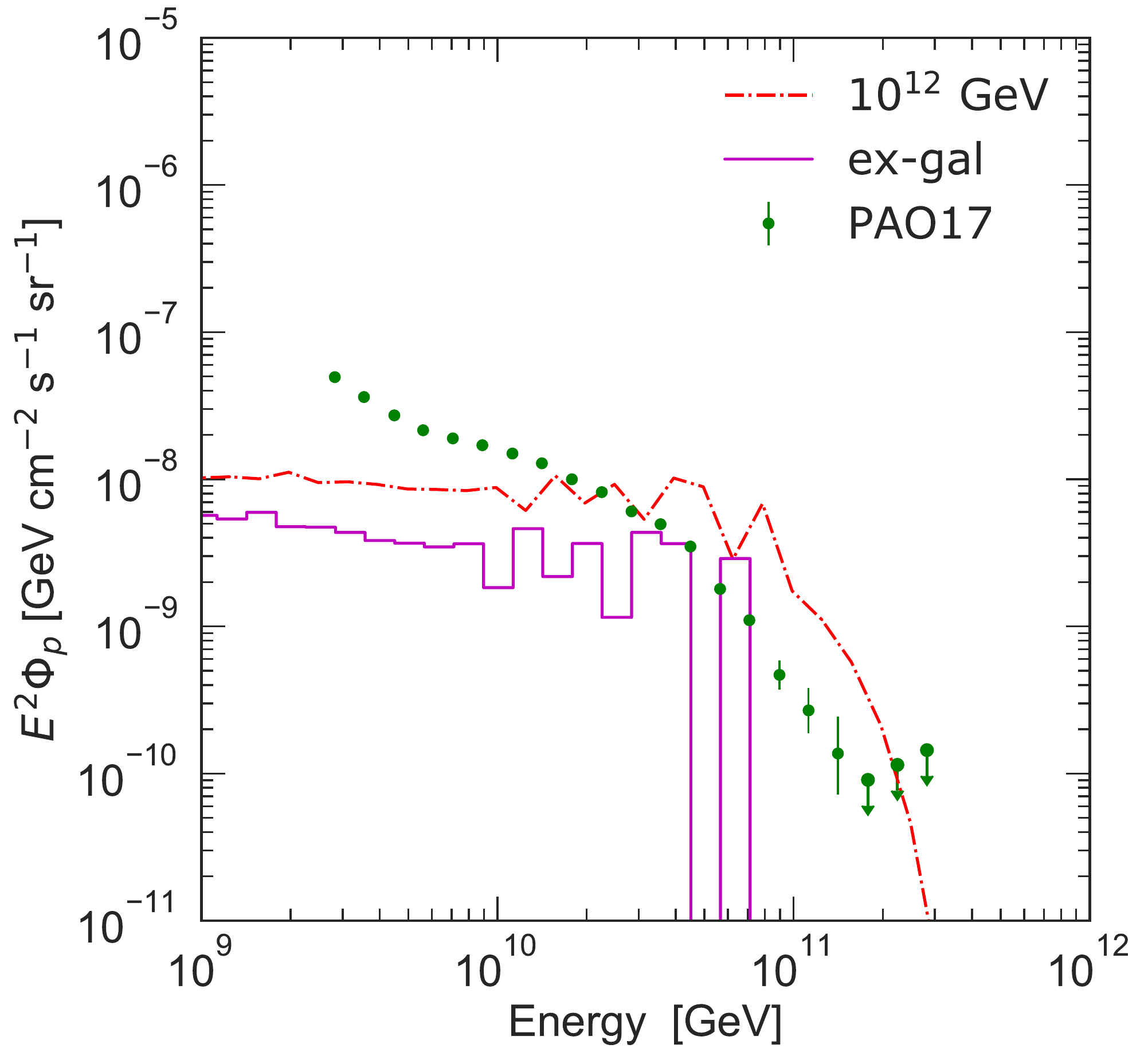}
   \includegraphics[width=7.5cm]{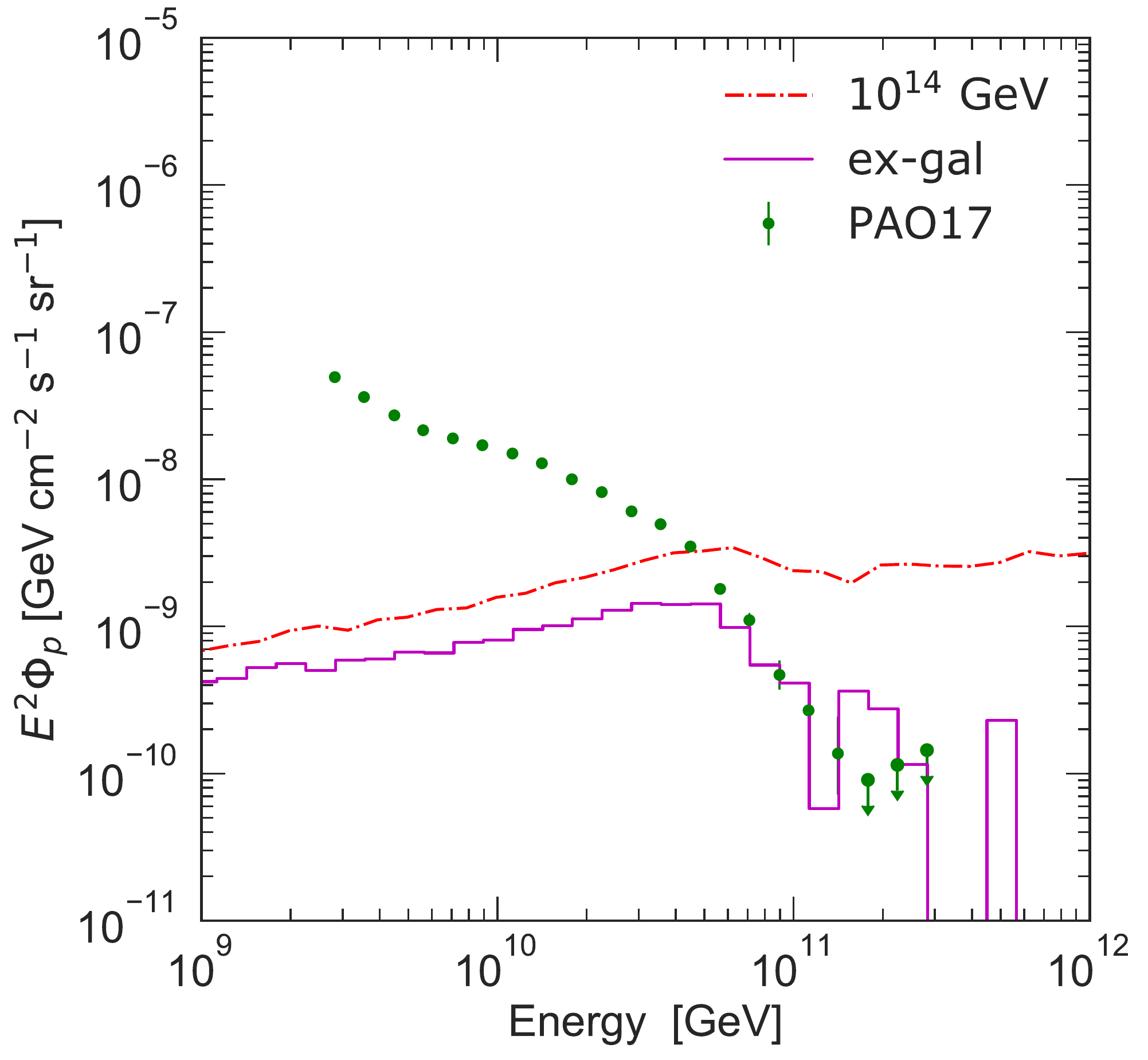}
   \includegraphics[width=7.5cm]{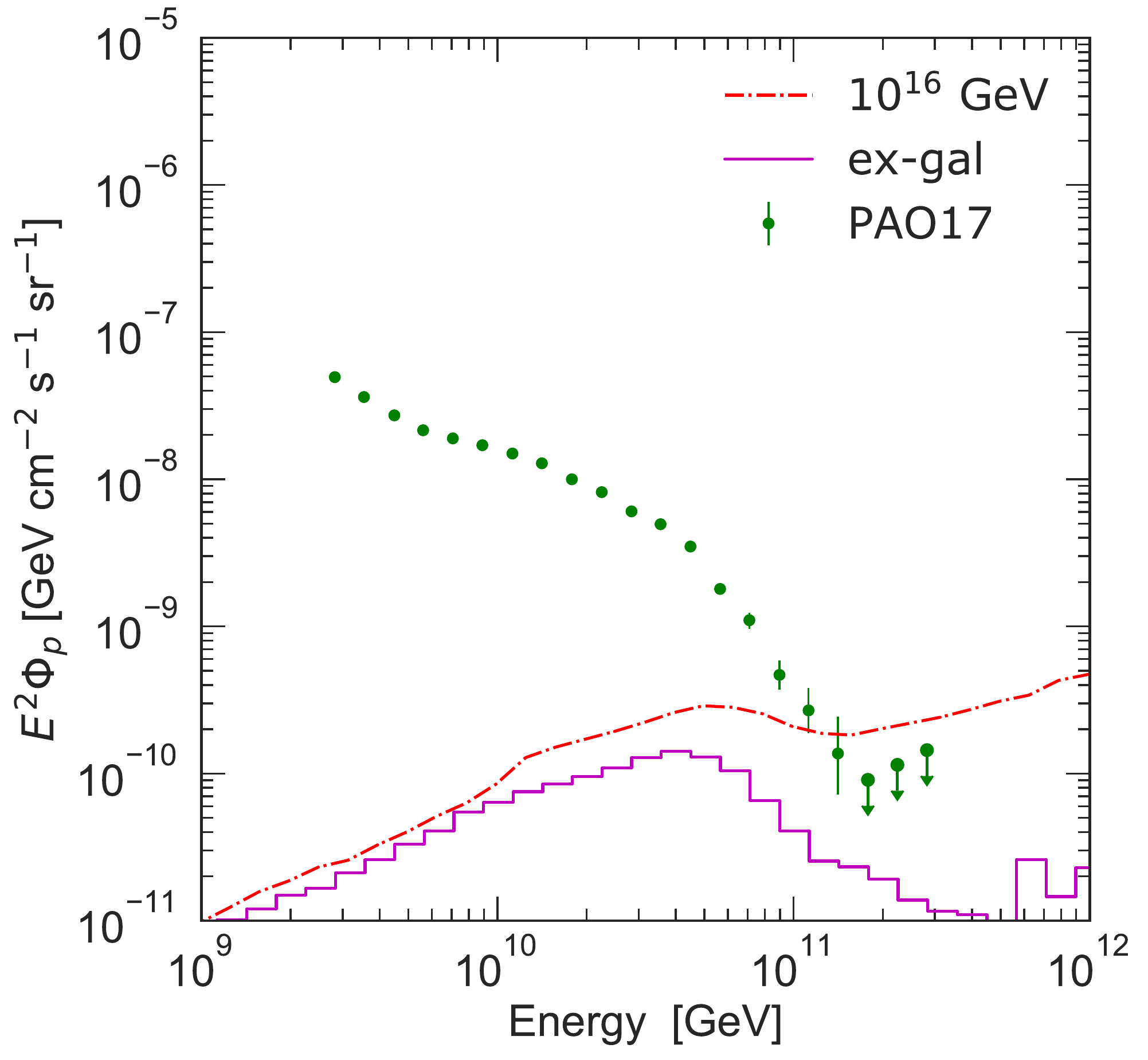}
   \caption{$p+\bar{p}$ fluxes due to dark matter decaying to
     $b\bar{b}$ where $m_{\rm dm}=10^{10}$, $10^{12}$, $10^{14}$, and
     $10^{16}$~GeV (from top to bottom, left to right), and the
     lifetime of dark matter is $10^{27}$~s.  Total flux (red
     dot-dashed) and extragalactic contribution (purple solid) are
     shown.  Data points correspond to the observed CR fluxes by
     PAO~\cite{Aab:2017njo}.}
  \label{fig:proton}
 \end{center}
\end{figure}

\begin{figure}[t!]
 \begin{center}
   \includegraphics[width=7.5cm]{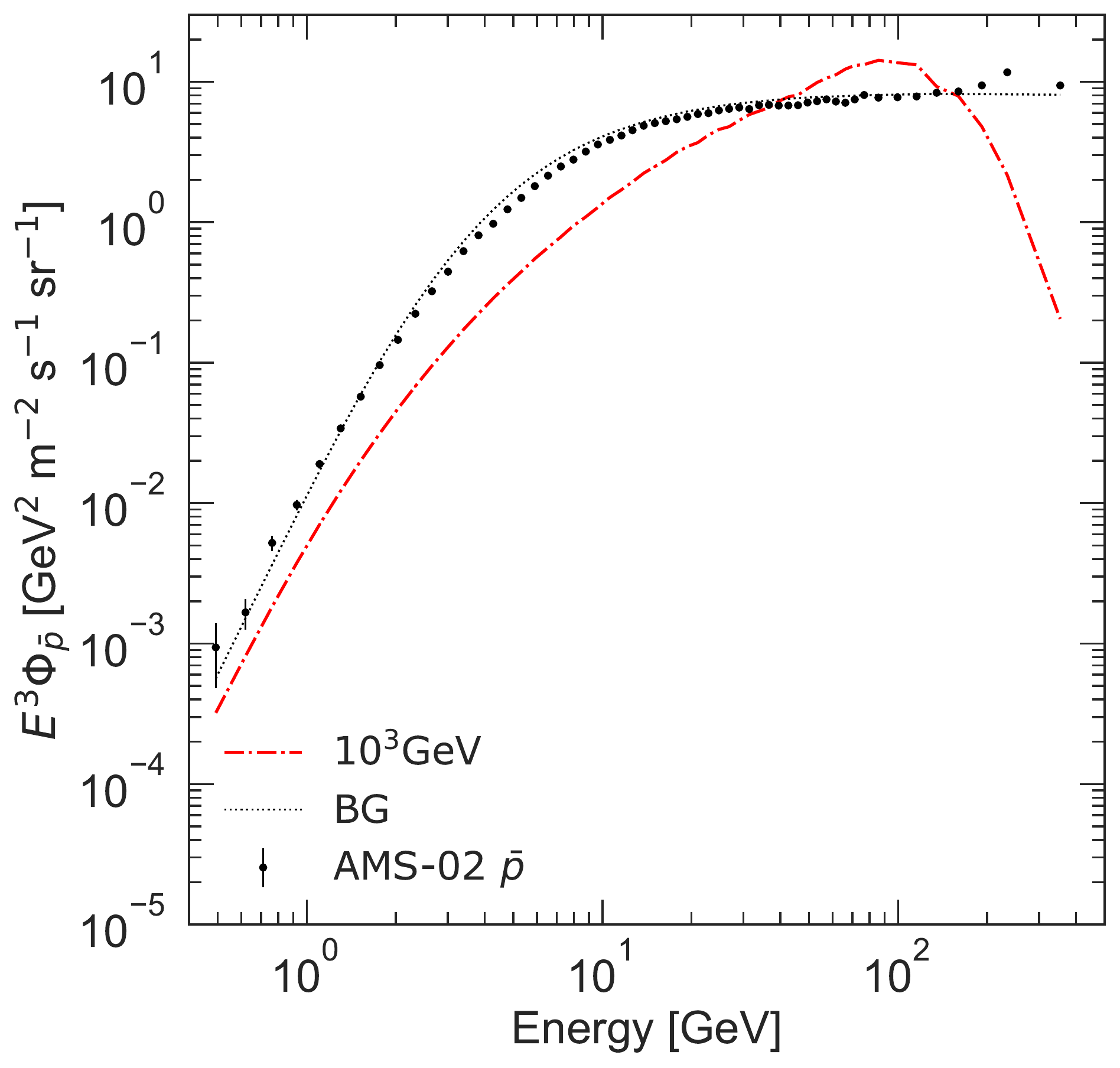}
   \includegraphics[width=7.5cm]{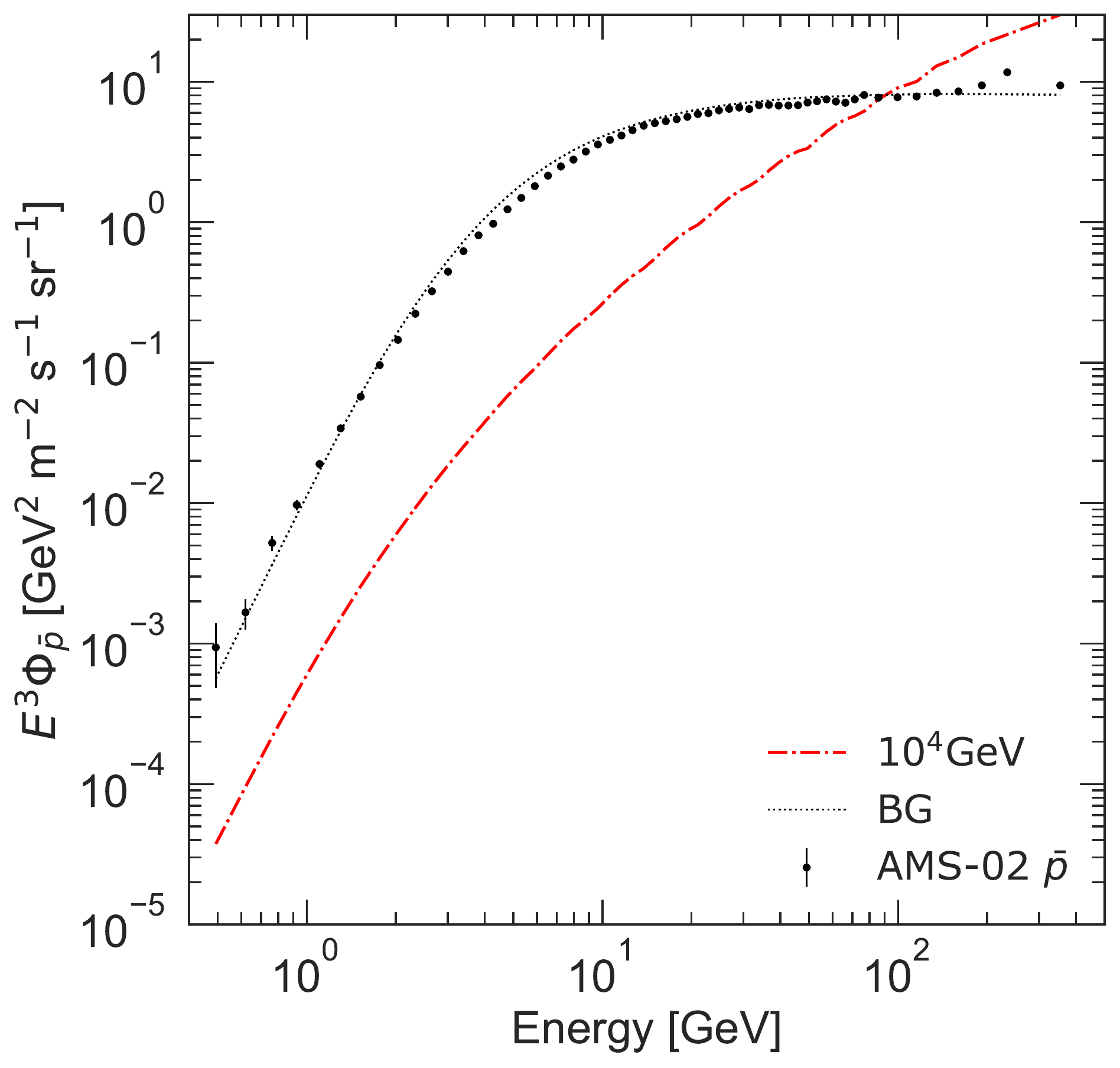}
   \caption{CR $\bar{p}$ spectra from decaying dark matter.  The
     propagation parameter setup used to determine the spectrum is
     shown in Tab.~\ref{tab:galprop_setup} and
     \ref{tab:helmod_setup}. The DM spectrum (red dot-dashed) is
     displayed for some particular DM mass values: $m_{\rm dm}=10^{3}$
     and $10^{4}$~GeV. The astrophysical background model (black
     dotted) reproduces the one found through a robust Markov chain
     Monte Carlo scan in Ref.~\cite{Boschini:2017fxq}. The data points
     are taken from AMS-02~\cite{Aguilar:2016kjl}}
  \label{fig:antiproton}
 \end{center}
\end{figure}

\begin{figure}[t!]
 \begin{center}
 \begin{tabular}{cc}
  \includegraphics[width=6.8cm]{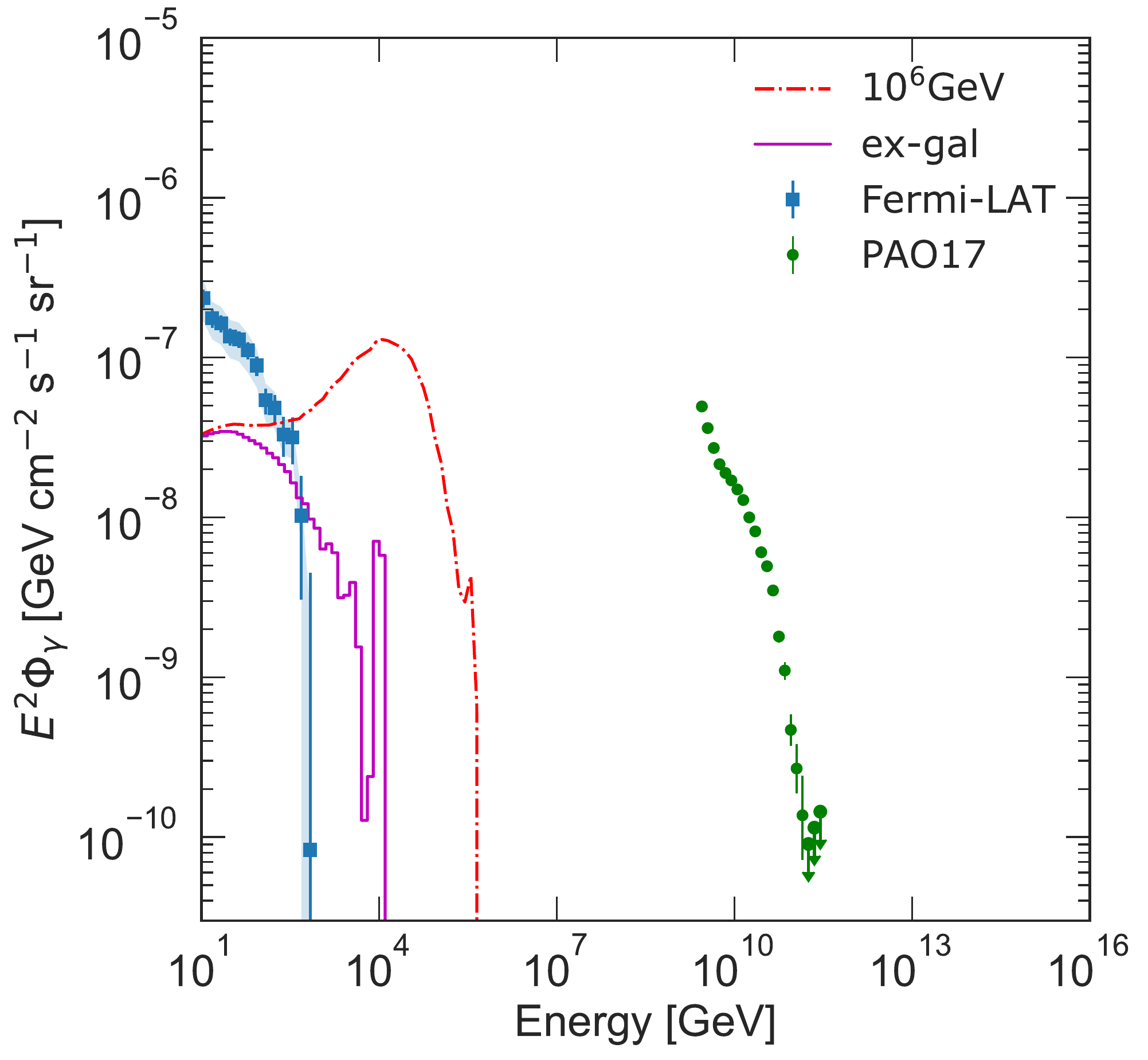} &
   \includegraphics[width=6.8cm]{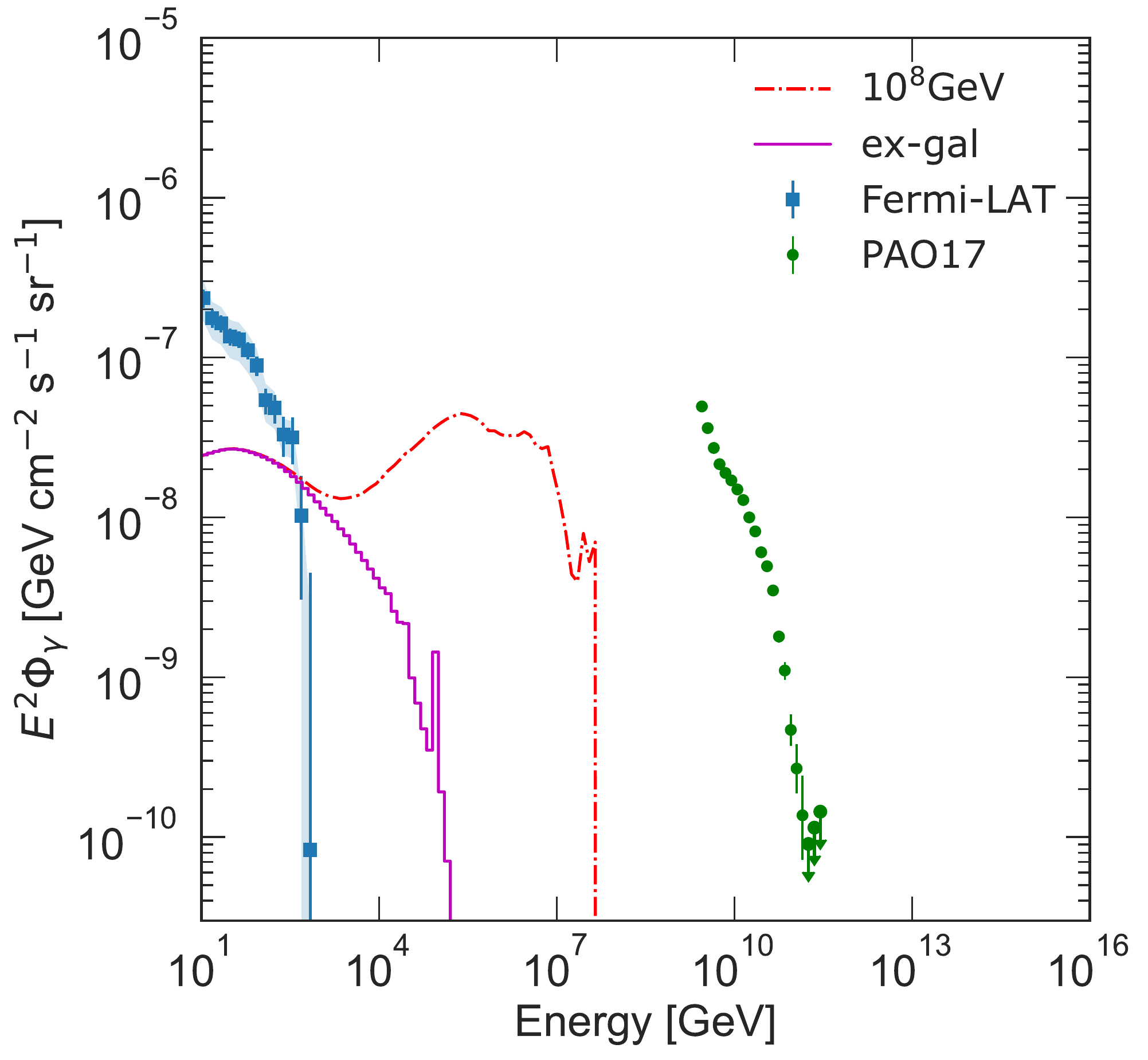}\\
   \includegraphics[width=6.8cm]{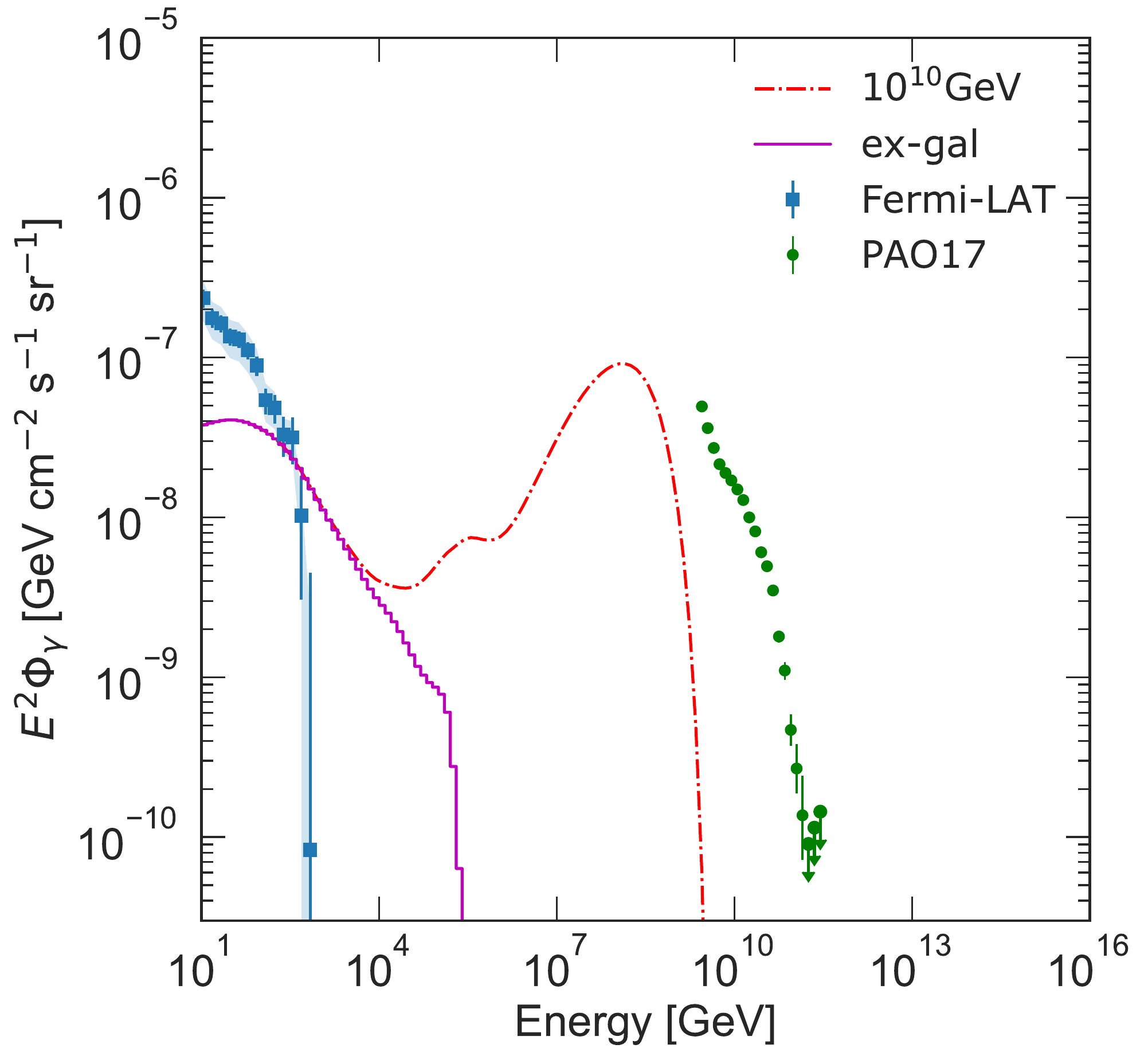}&
   \includegraphics[width=6.8cm]{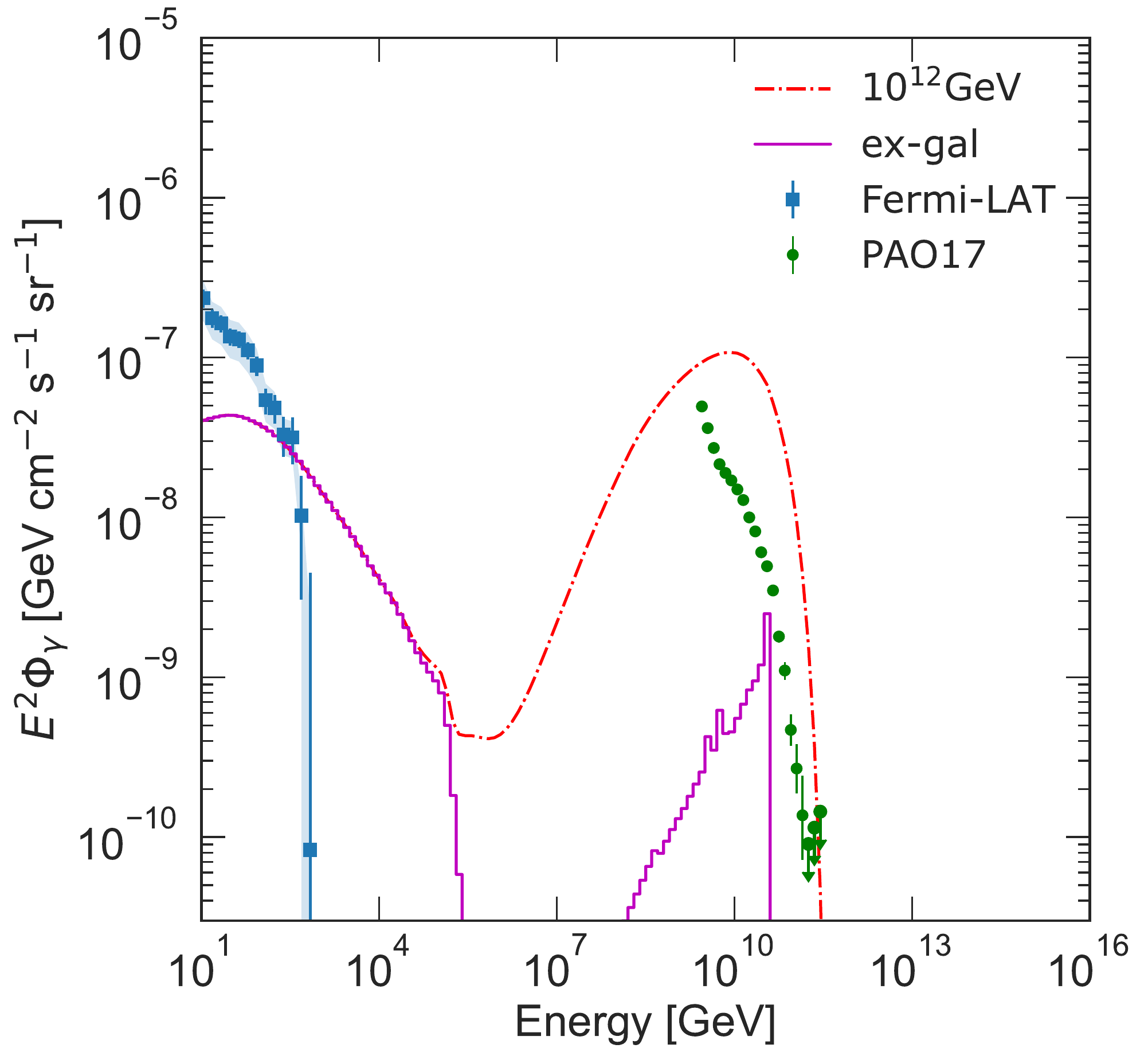}\\
   \includegraphics[width=6.8cm]{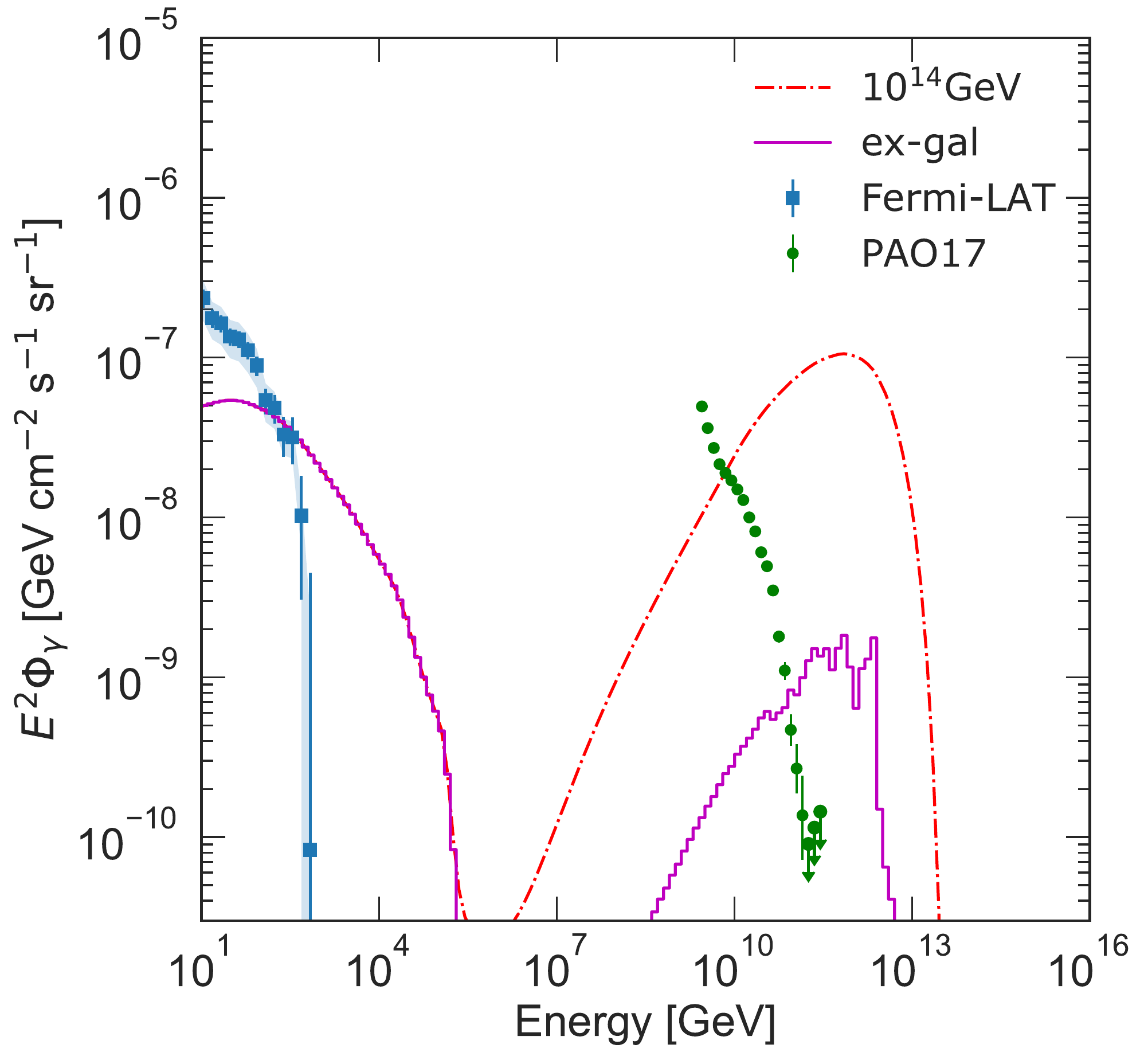}&
   \includegraphics[width=6.8cm]{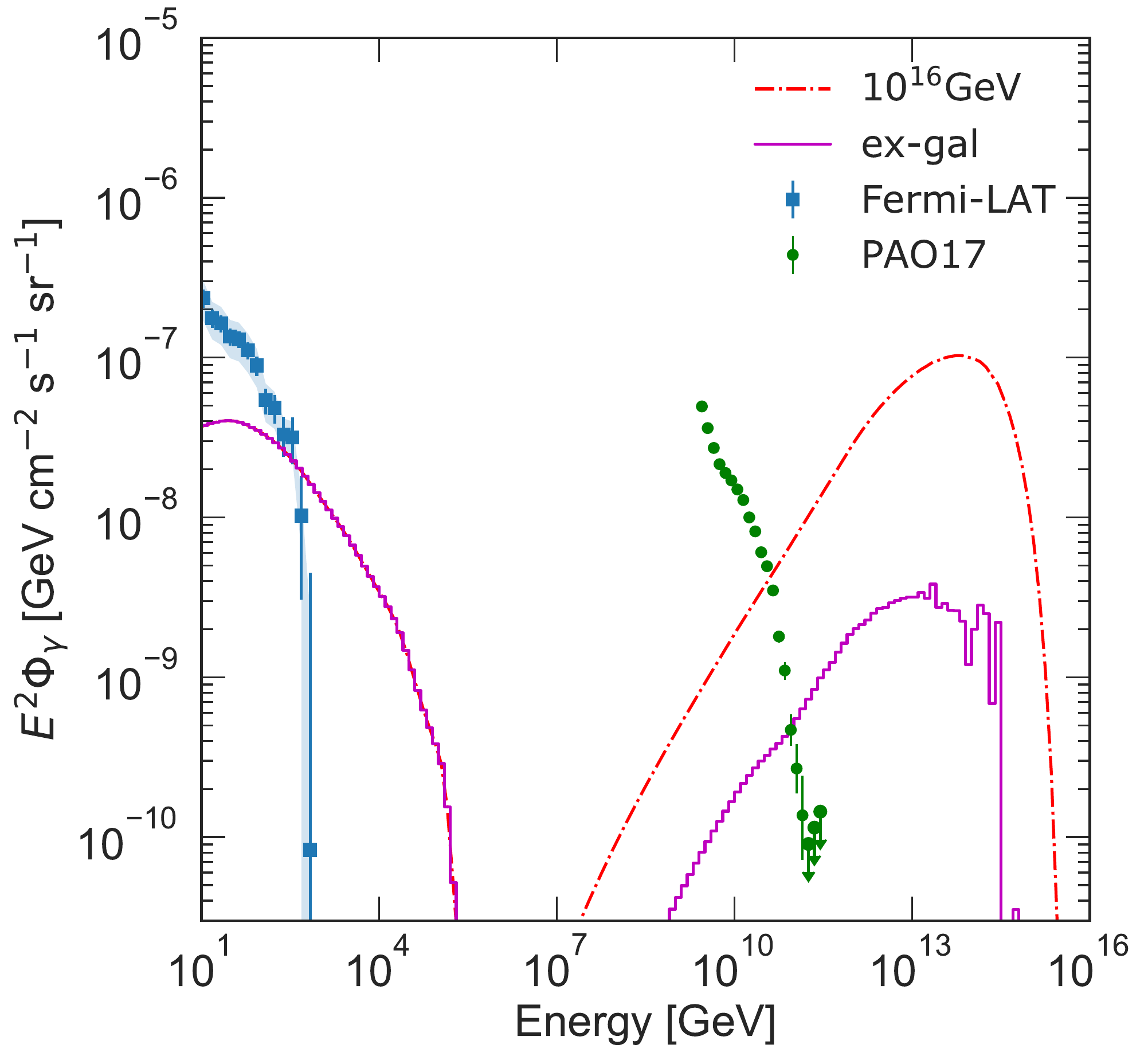}
  \end{tabular}
\caption{CR $\gamma$ spectra from decaying DM particles into the
  $\bar{b}b$ channel. See text for descriptions of the modelling
  assumptions. Components shown in each panel follow the same
  conventions as in Fig.~\ref{fig:proton}. Shown are DM masses of
  $m_{\rm dm}=10^6$, $10^8$, $10^{10}$, $10^{12}$, $10^{14}$ and
  $10^{16}$ GeV. Photon spectral measurements are taken from
  Fermi-LAT~\cite{Ackermann:2014usa} and PAO~\cite{Aab:2017njo}.}
  \label{fig:gamma}
 \end{center}
\end{figure}

\begin{figure}[t!]
  \begin{center}
  \begin{tabular}{cc}
\includegraphics[width=6.8cm]{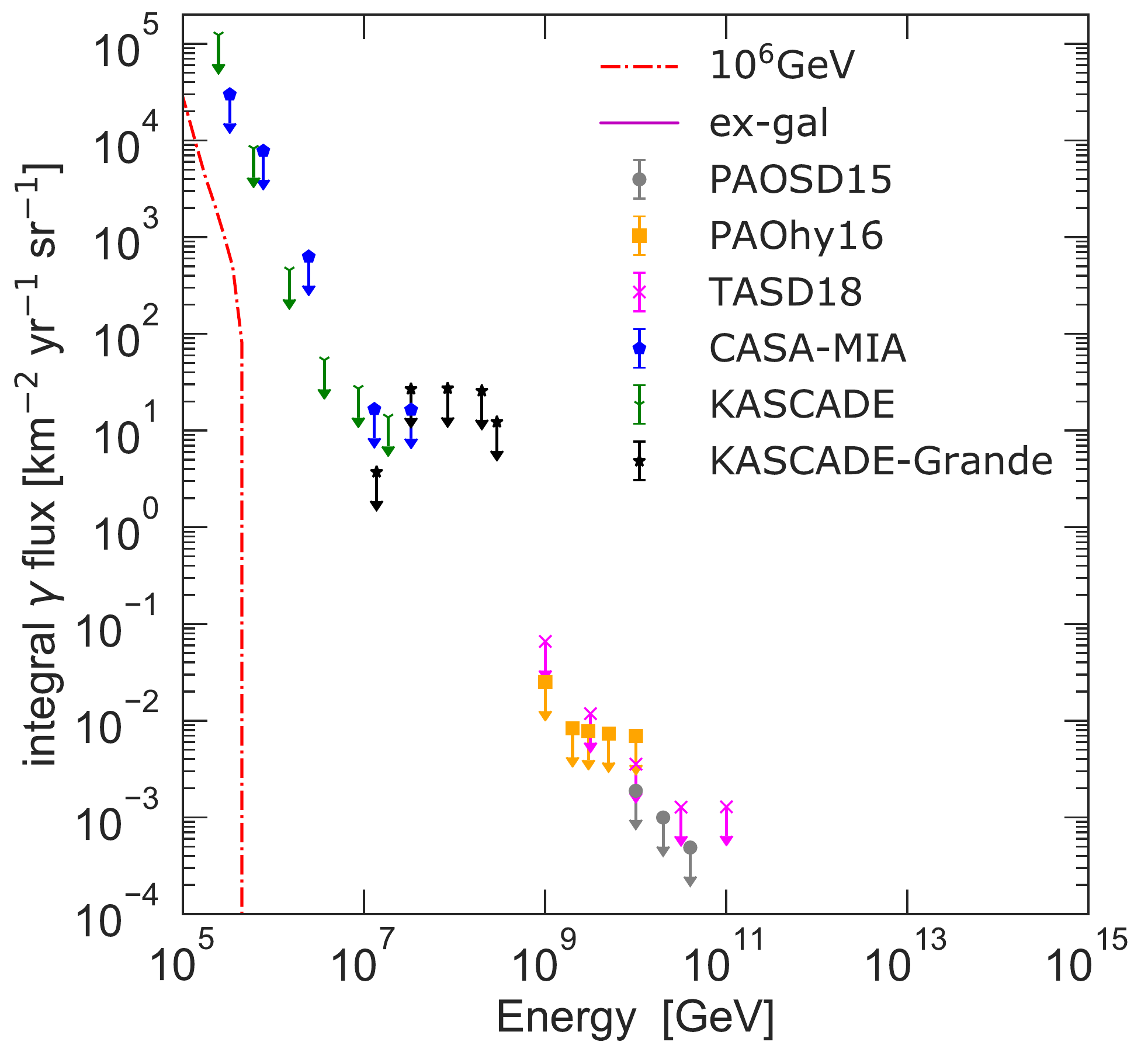}&
    \includegraphics[width=6.8cm]{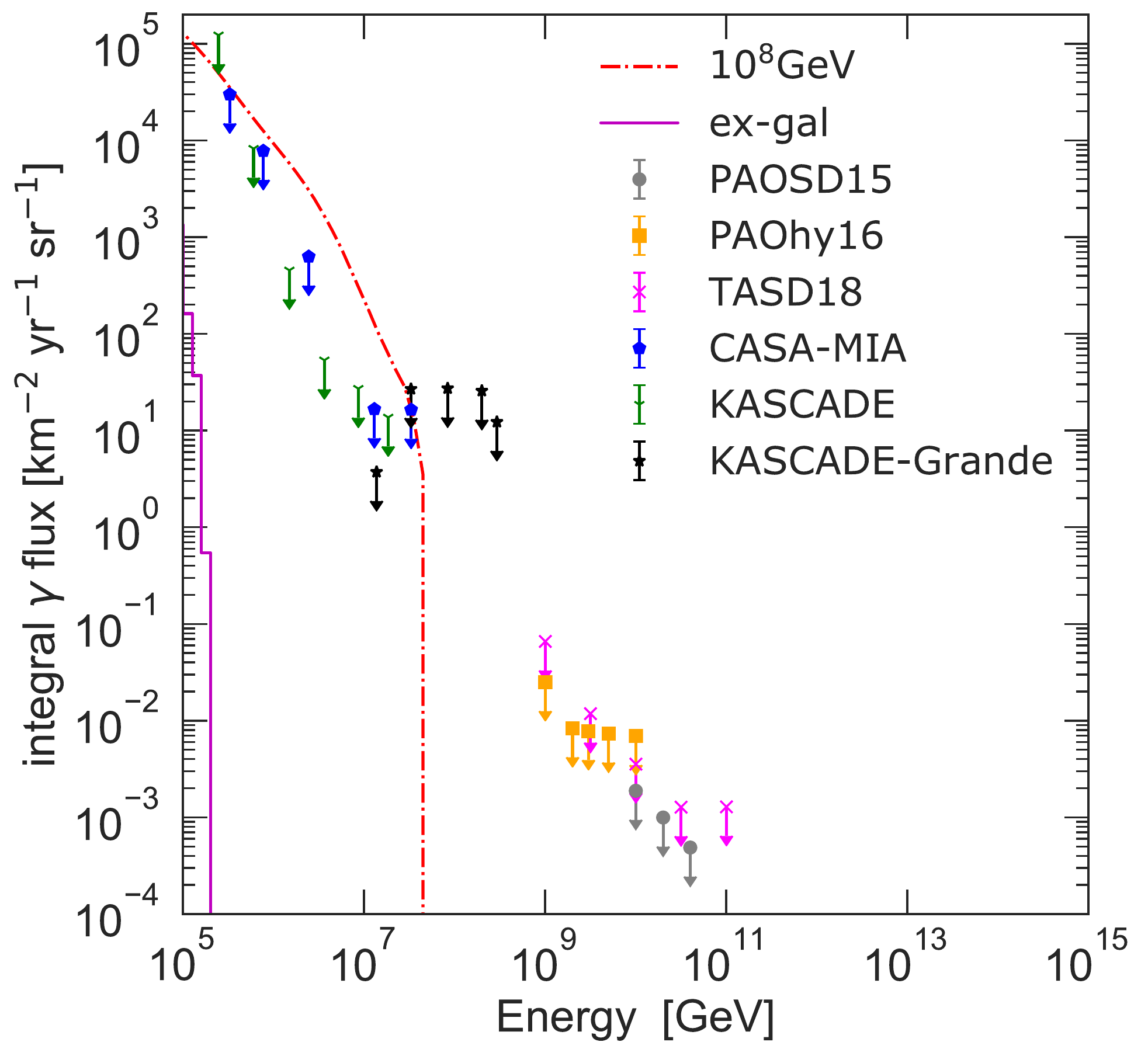}\\
    \includegraphics[width=6.8cm]{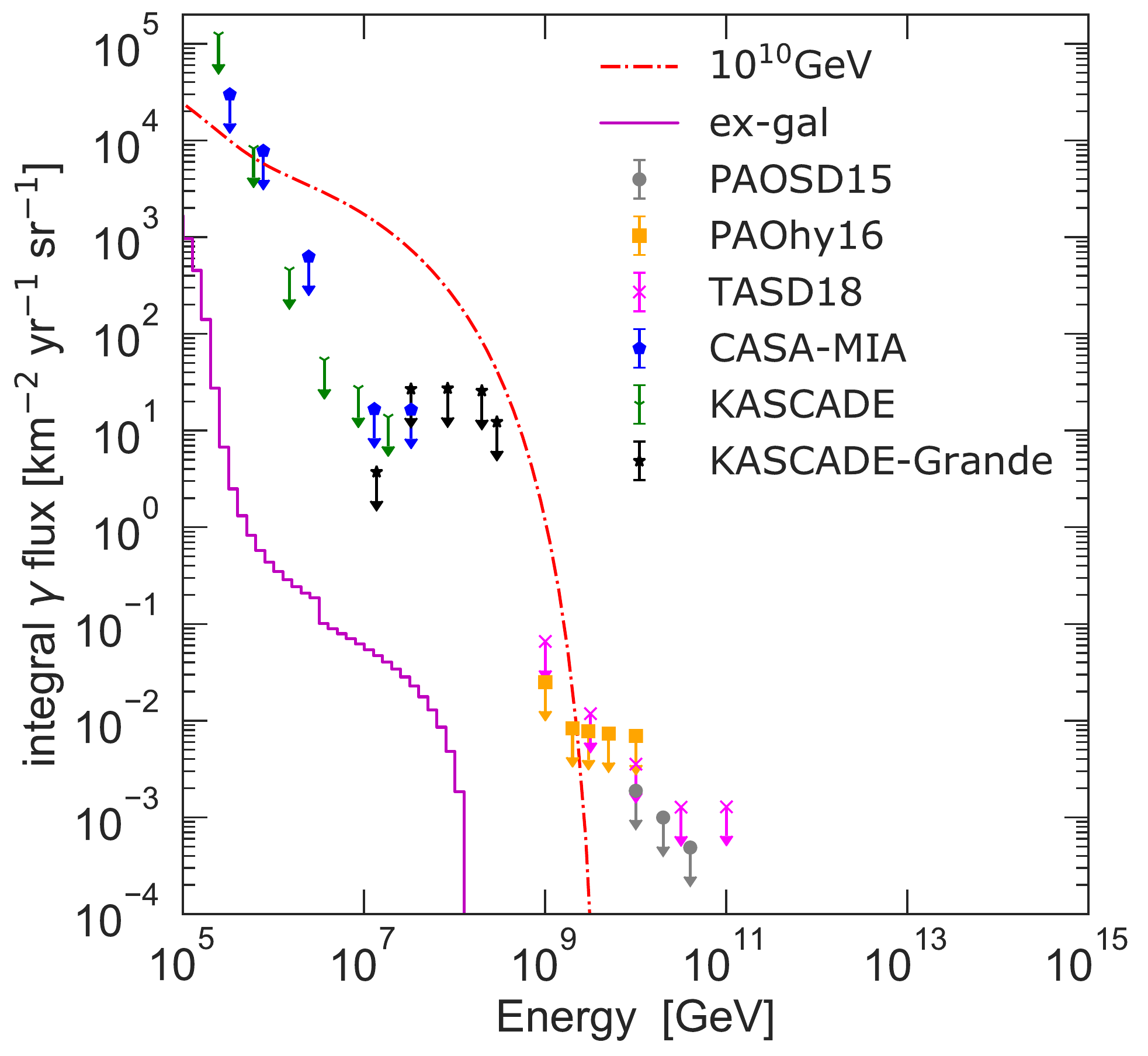}&
    \includegraphics[width=6.8cm]{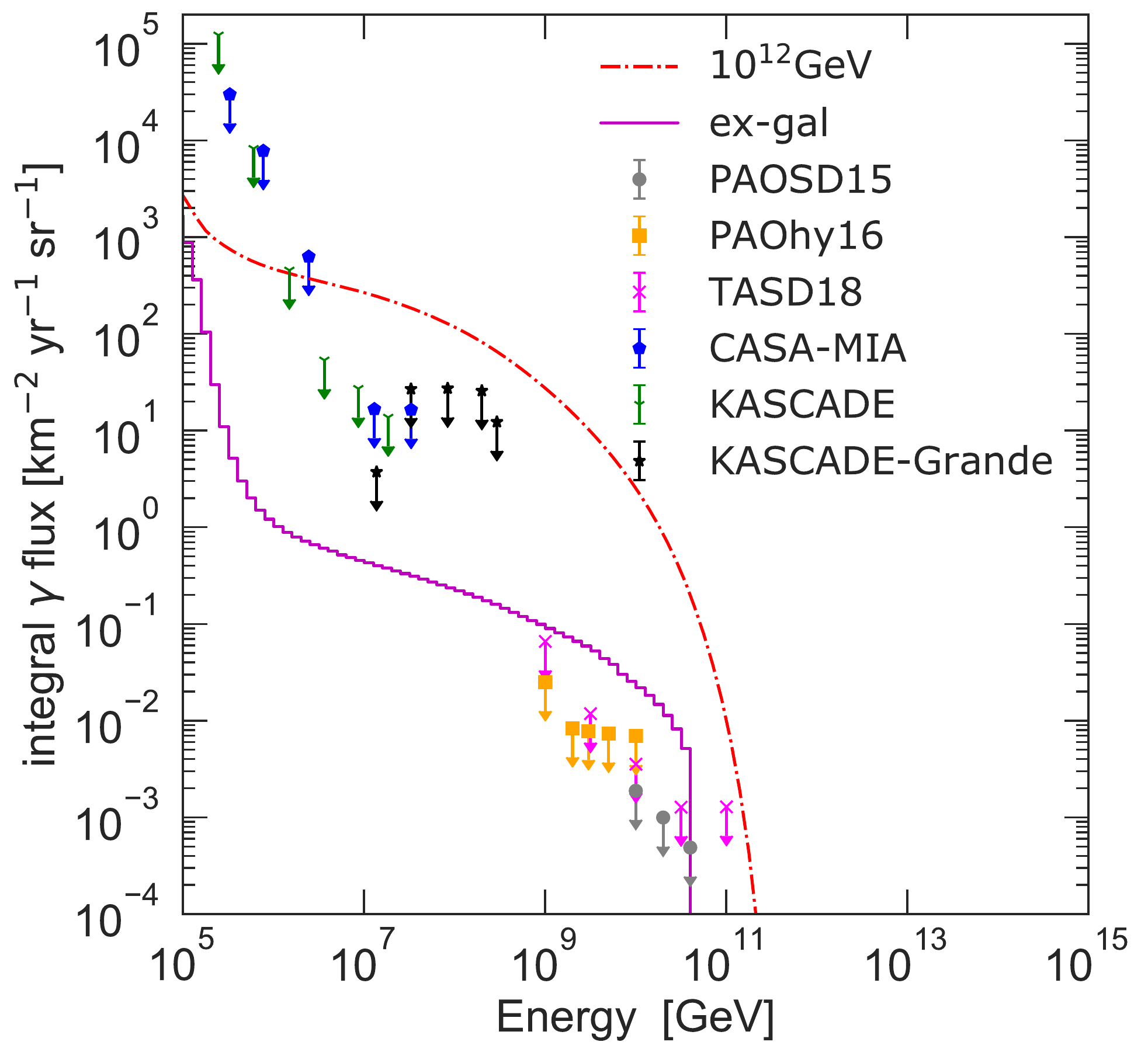}\\
    \includegraphics[width=6.8cm]{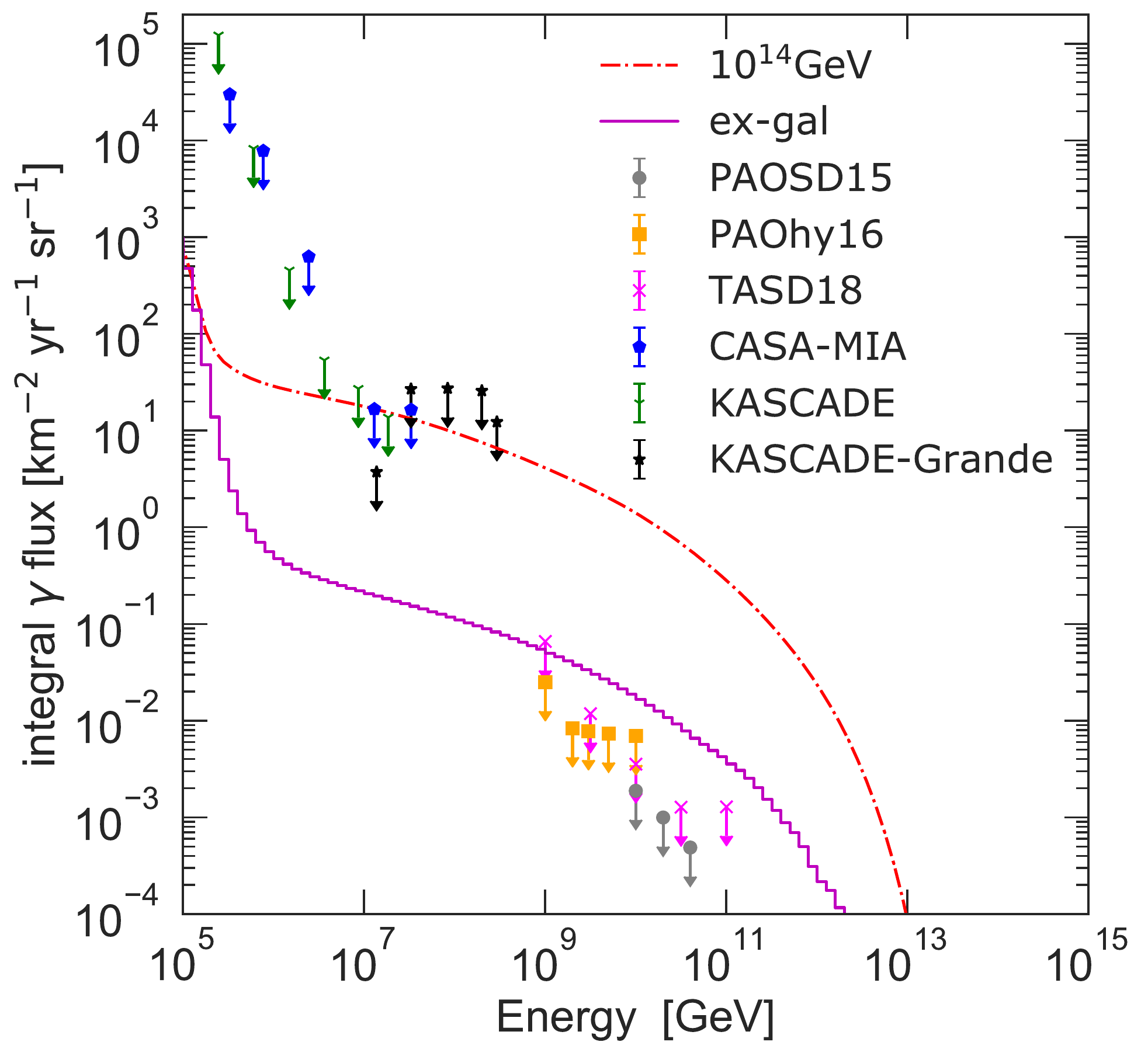}&
    \includegraphics[width=6.8cm]{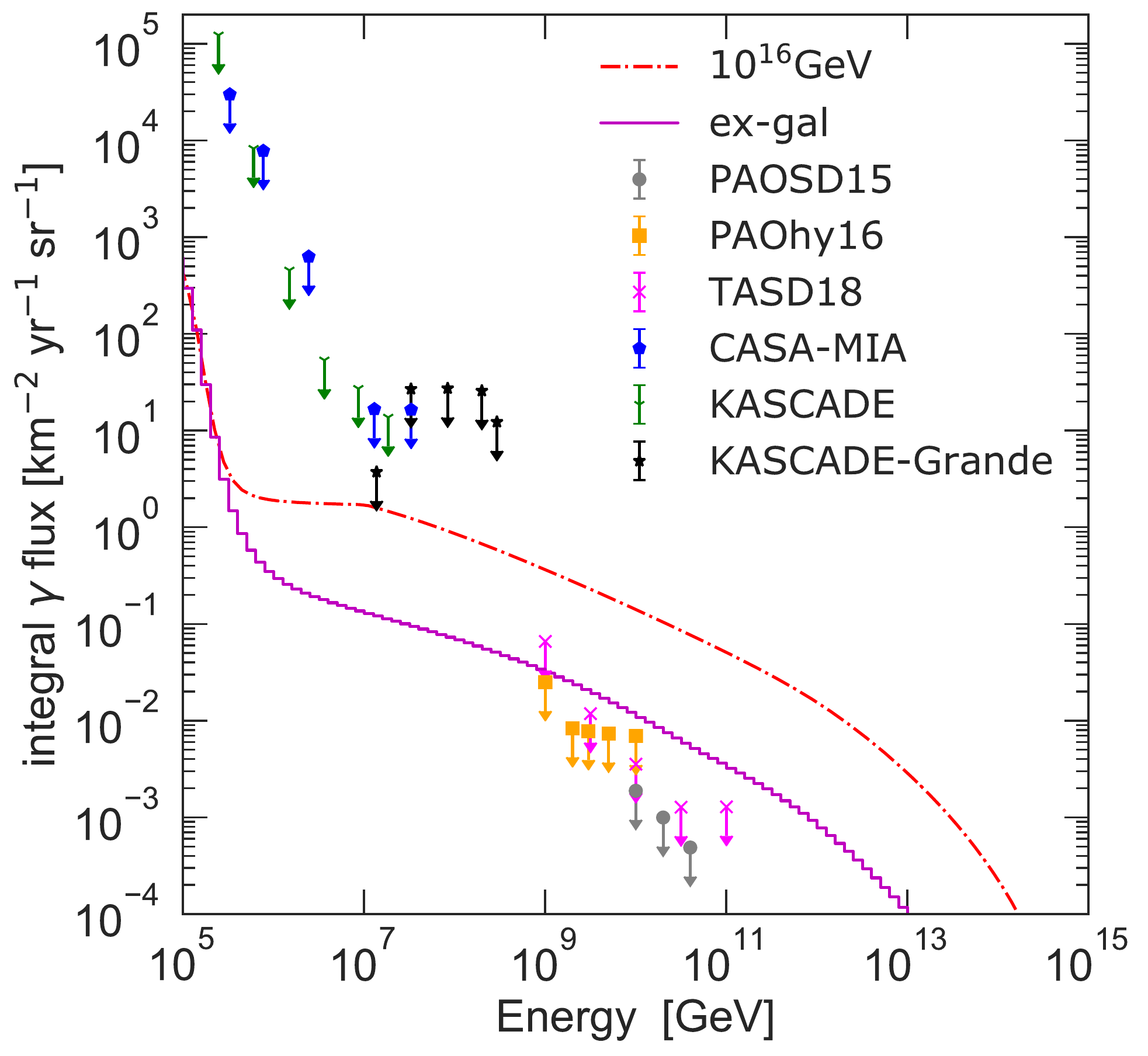}
  \end{tabular}
     \caption{Integrated $\gamma$ fluxes. Modelling parameters are
       taken to be the same as in Fig.\,\ref{fig:gamma}. Upper bounds
       from the observations are given by
       CASA-MIA~\cite{Chantell:1997gs}, KASCADE,
       KASCADE-Grande~\cite{Apel:2017ocm},
       PAO~\cite{Aab:2015bza,Aab:2016agp} and
       TA~\cite{Abbasi:2018ywn}.}
  \label{fig:integgamma}
 \end{center}
\end{figure}

\begin{figure}[t!]
 \begin{center}
 \begin{tabular}{cc}
 \includegraphics[width=6.8cm]{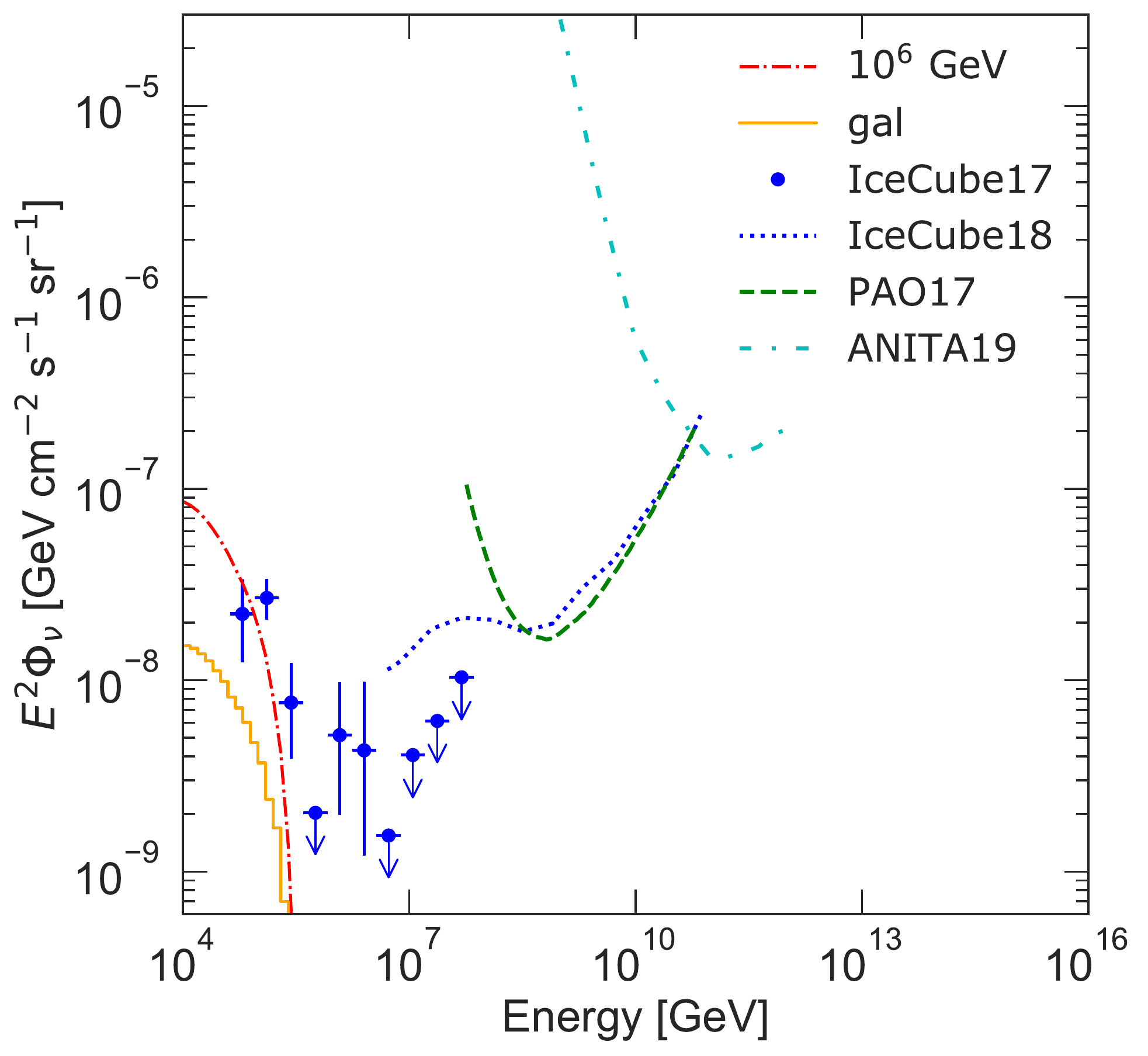}&
   \includegraphics[width=6.8cm]{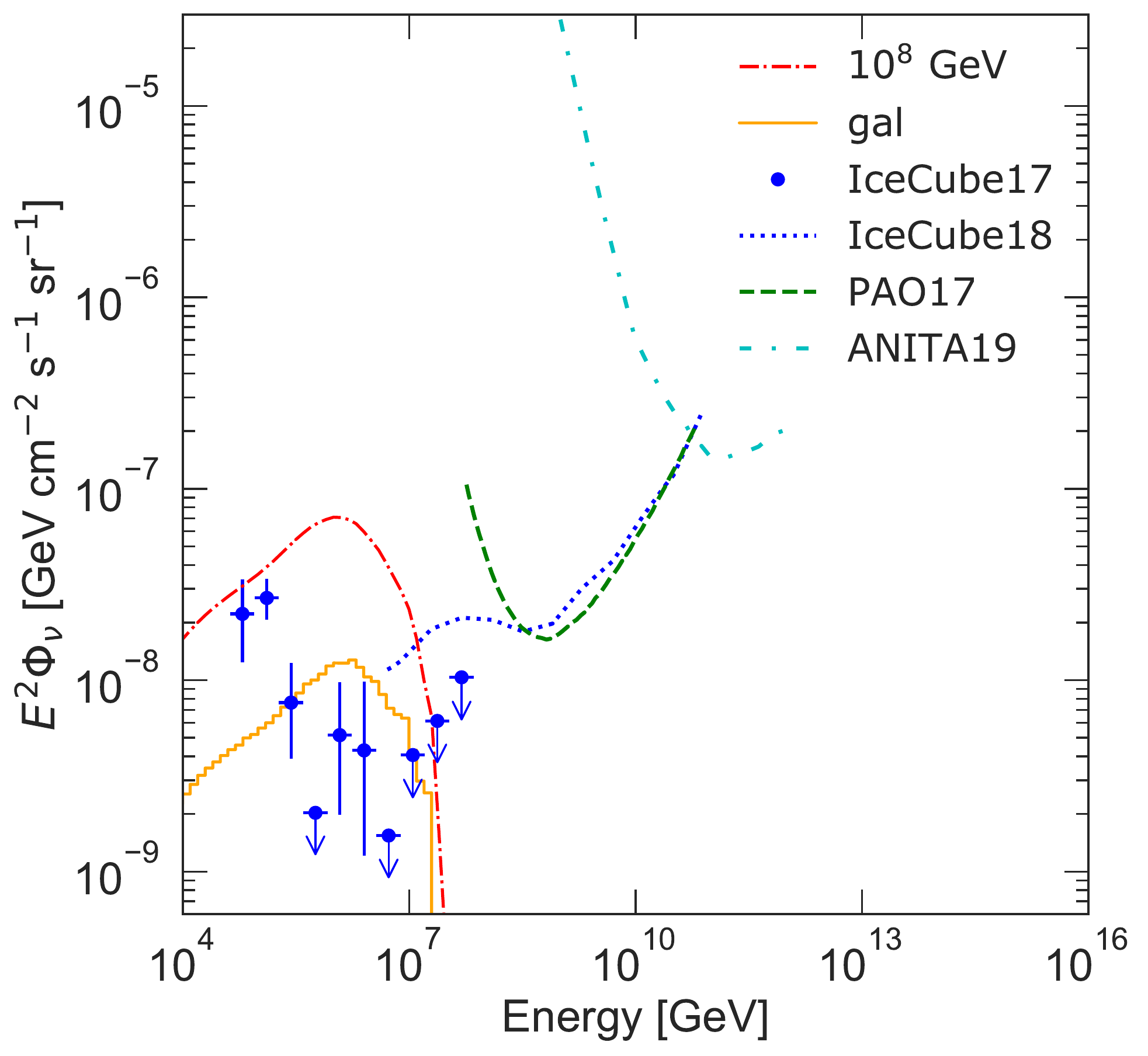}\\
   \includegraphics[width=6.8cm]{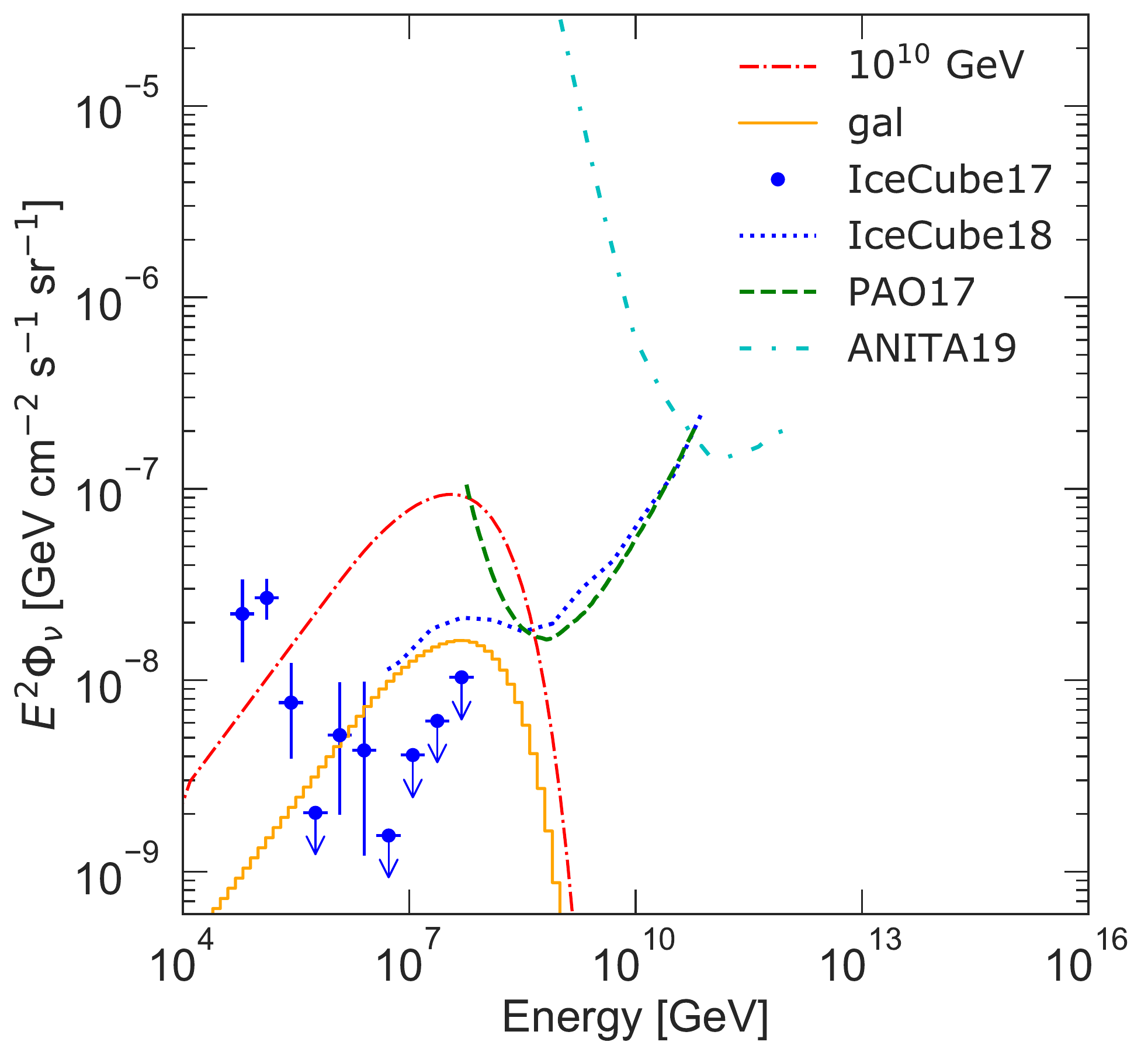}&
   \includegraphics[width=6.8cm]{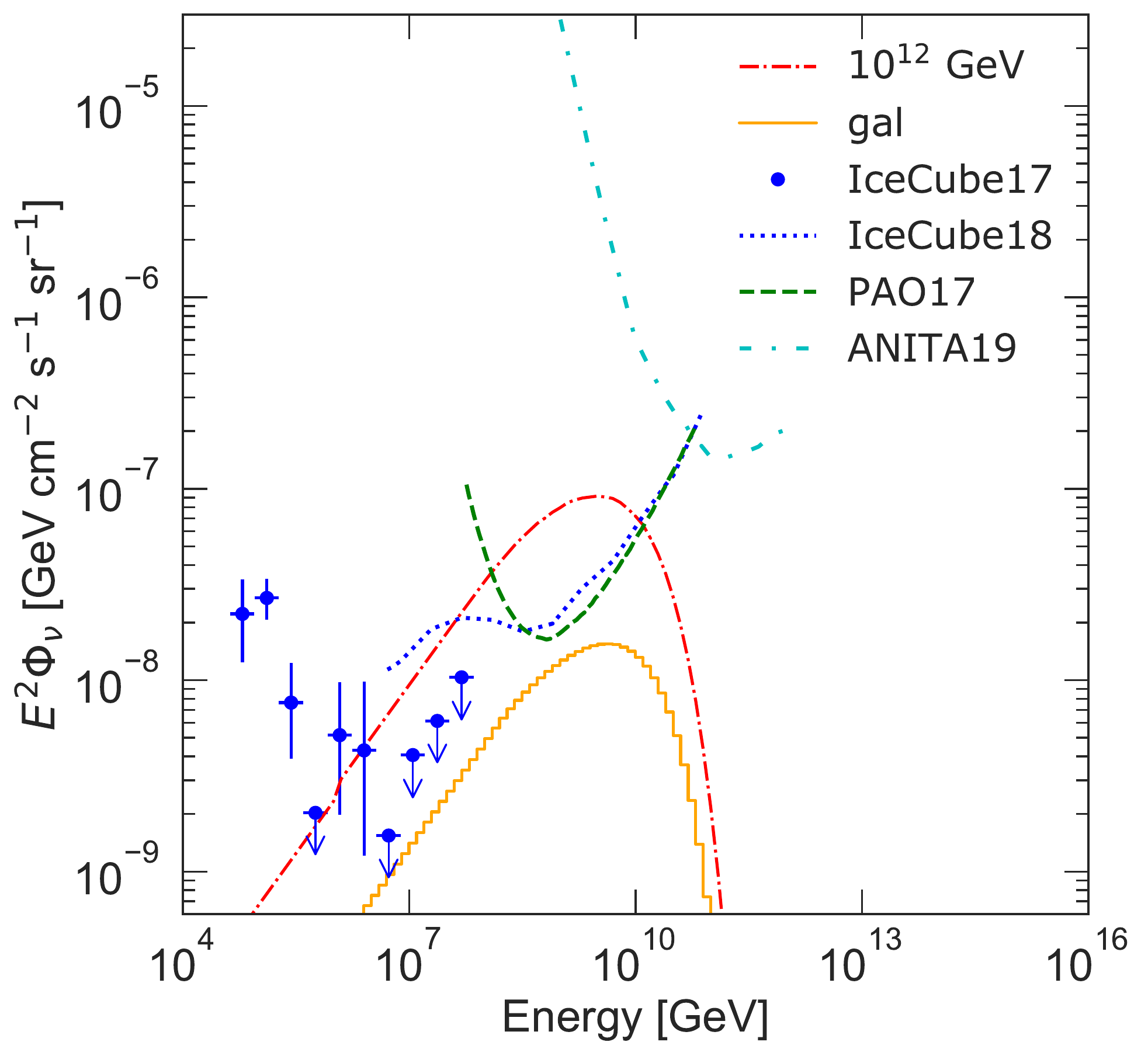}\\
   \includegraphics[width=6.8cm]{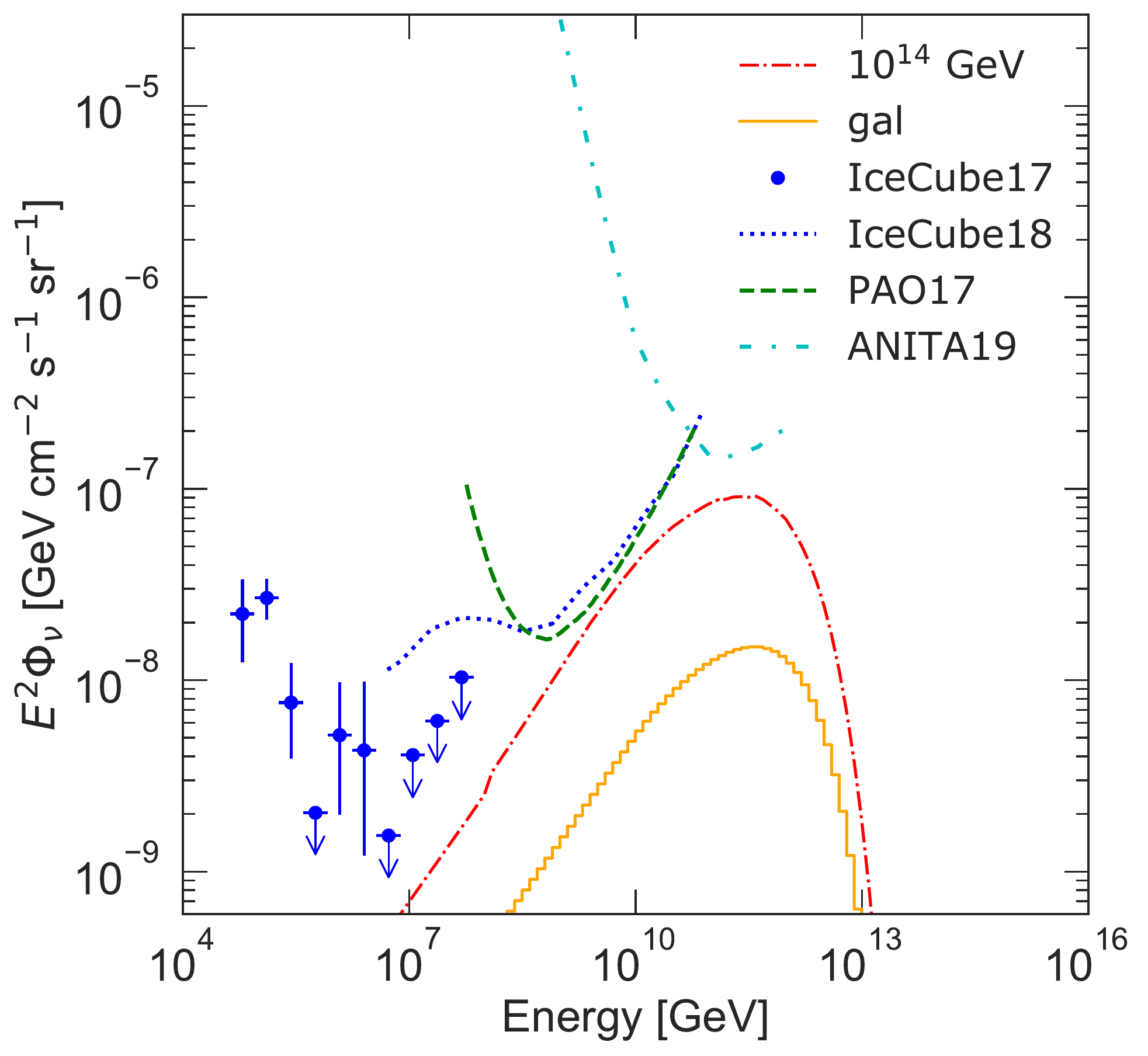}&
   \includegraphics[width=6.8cm]{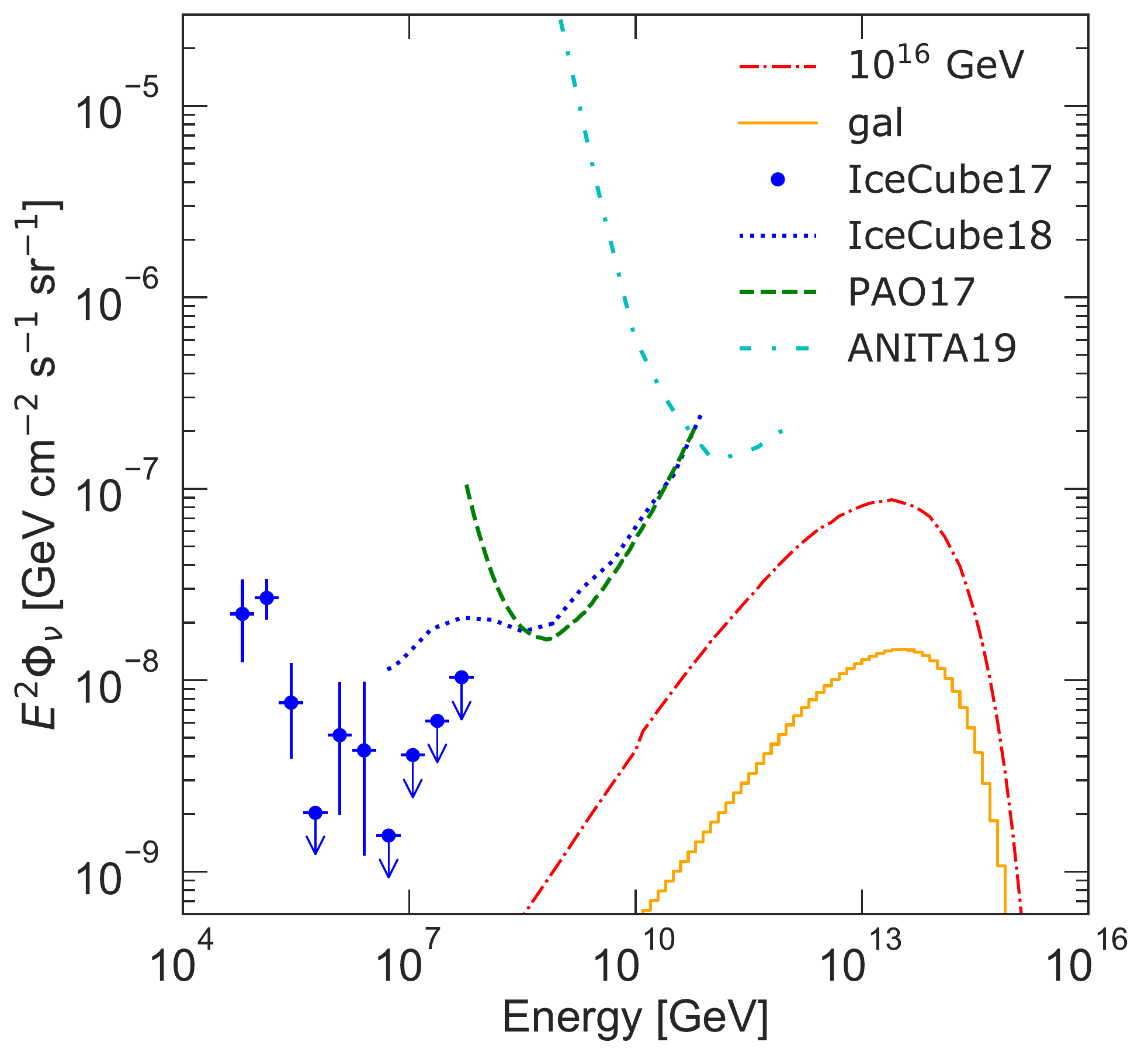}     
 \end{tabular}
   \caption{$\nu+\bar{\nu}$ fluxes. Modelling parameters are taken to
     be the same as in Fig.\,\ref{fig:proton}. The spectrum obtained
     from propagation in the Galactic region (yellow solid) is plotted
     in addition to the total spectrum (red dot-dashed).  Data points
     are from IceCube~\cite{Kopper:2017zzm}, and the others are upper
     bounds by IceCube~\cite{Aartsen:2016ngq}, PAO~\cite{Aab:2017njo},
     and ANITA~\cite{Gorham:2019guw}.}
  \label{fig:neutrino}
 \end{center}
\end{figure}

In this section we present our procedure to derive conservative
constraints on the DM lifetime. Except for a few cases detailed below,
we do not subtract background/foreground models and only require that
any putative DM signal does not overshoot the observed CR flux at any
given energy bin. In practice, this means that our lower limits on the
DM lifetime are calculated by varying the dark matter lifetime until
the observed CR flux is saturated. However, when using $\gamma$-rays
(Fermi-LAT) and $\bar{p}$ (AMS-02) data, we will run our lower limits
pipeline by taking into account the respective background/foreground
models. This is because our understating of the astrophysical
background for these particular channels has increasingly improved
recently, and thus, we can be less conservative when using these data
sets.

For the two exceptions mentioned above we use the F-test to compute
the 95\% CL lower limits. This is done by comparing the null model
(background-only hypothesis), with the alternate model (background
plus DM hypothesis), where the DM flux norm is fixed to a specific
value. This value is then changed until the difference between
$\chi^2_{\rm Null}$ and $\chi^2_{\rm Alternate}$ cannot be explained
by the loss of a degree of freedom within a 95\% confidence. In
particular,
\begin{equation}\label{eq:lmfit}
  F(l_{\rm fixed},N-l_{\rm Null})   =
  \left(\frac{\chi^2_{\rm Alternate}}{\chi^2_{\rm Null}}-1 \right)
  \frac{N-l_{\rm Null}}{l_{\rm fixed}},
\end{equation}
where $l_{\rm Null}$ is the number of parameters of the null model,
$N$ is the number of data points and $l_{\rm fixed}$ is the difference
of number of parameters between the null and alternate hypotheses. We
solve Eq.\,\eqref{eq:lmfit} with the non-linear least-squares
minimization package
\texttt{lmfit}.\footnote{\url{https://lmfit.github.io/lmfit-py/intro.html}}

\subsection{Cosmic ray fluxes}
\label{sec:fluxes}

Using Eq.\,\eqref{eq:spectraatsource} and the propagation methods
explained in the previous section we compute predicted CR fluxes at
Earth originating from DM masses larger than $10^3$ GeV. Several
particular examples of our predictions along with their respective
data sets (that will be used to impose constraints) are shown
below. Specifically, here we show observable fluxes for CR $p$,
$\bar{p}$, $\gamma$-rays, and $\nu$. All the CR spectra shown in
this section assume a DM lifetime of $10^{27}$ s.

Figure\,\ref{fig:proton} displays the $p+\bar{p}$ fluxes for DM masses
of $10^{10}$, $10^{12}$, $10^{14}$, and $10^{16}$~GeV (from top to
bottom, left to right). As it can be seen, the Galactic components are
comparable to the extragalactic ones for $m_{\rm dm}\lesssim
10^{11}\,{\rm GeV}$, however the later become dominant for larger DM
masses. We anticipate that more stringent bounds on DM lifetime will
be obtained by using the predicted Galactic CR spectra.  Furthermore,
while the extragalactic contributions are suppressed for $m_{\rm
  dm}\gtrsim 10^{11}\,{\rm GeV}$, its overall intensity remains
unchanged up to $m_{\rm dm}\sim 10^{11}\,{\rm GeV}$. This behavior is
a result of the GZK effect.  Namely, $p$ ($\bar{p}$) lose their
energies due to photo-pion production process which is relevant for
$p$ energy over $10^{11}\,{\rm GeV}$.  Then part of that lost
energy is converted into pions, whose decay products emit a given
amount of $\gamma$, $e^\pm$ and $\nu$, $\bar{\nu}$.  Although their
fluxes are suppressed for $E\gtrsim 10^{11}\,{\rm GeV}$, these are
nonetheless comparable to the observed CR fluxes at Earth. Thus,
models of new physics predicting DM particles with $\tau_{\rm
  dm}\lesssim 10^{27}\,{\rm s}$ and $m_{\rm dm}\gtrsim 10^{10}\,{\rm
  GeV}$ are expected to be constrained by observations.

We show the $\bar{p}$ spectra for $m_{\rm dm}=10^{3}$ and $10^4$\,GeV
in Fig.\,\ref{fig:antiproton}. In this figure, the astrophysical
background is also shown. As explained in the previous section, the
astrophysical background used in this work reproduces the one explored
in Ref.\,\cite{Boschini:2017fxq}.\footnote{The antiproton background
  computed with the GALPROP-HelMod method in
  Ref.~\cite{Boschini:2017fxq} slightly overpredicts the AMS-02
  measurements at ~10 GeV. However, no such discrepancy is observed
  when the predictions are compared to PAMELA
  data~\cite{Boschini:2017fxq}. It should be mentioned that the MCMC
  scan procedure performed in that study included antiproton data from
  PAMELA, AMS-02 and BESS-Polar II. So systematic uncertainties in
  that energy range explain any apparent discrepancy between
  background model predictions and observations. In our work we
  confirm such results and set out to impose constraints on decaying
  DM particles.} In this case we find that the extragalactic flux
spectra is negligibly small for this energy range. In addition, it can
be noticed that the $\bar{p}$ flux gets suppressed as the DM mass
increases. It will be shown in the next section, that the resulting
constraints for this channel (using AMS-02 data) are stringent around
$m_{\rm dm}\sim 1\,{\rm TeV}$ but become weaker for larger DM
masses. Using the same propagation parameter setup as for other CR
species, the $e^+$ spectra is computed. It turns out that the flux is
much smaller than the AMS-02 $e^+$ data for $\tau_{\rm
  dm}=10^{27}\,{\rm s}$ and that it is suppressed when the DM mass
gets large. We have found the constraints from the AMS-02 $e^+$ data
is irrelevant.

Figure\,\ref{fig:gamma} shows $\gamma$ fluxes for the same mass values
assumed in Fig.\,\ref{fig:proton}. The spectral bump seen in the high
energy regime corresponds to the contribution from the Galactic
component. We find that the $\gamma$ rays due to the ICS and
bremsstrahlung in the Galaxy are subdominant in the total flux. The
extragalactic component, on the other hand, exhibits two spectral
peaks; one at low energies and another one at high energies. The
former originates in the cascades from prompt DM decays, while the
later arises from electromagnetic cascades of $\gamma$ and $e^\pm$
coming from photo-hadronic processes. In all the panels we observe an
energy range ($10^5\,{\rm GeV} \lesssim E \lesssim 10^{10}\,{\rm
  GeV}$) where the emission of $\gamma$ is suppressed. This is because
the PP process is so effective that photons with these energies lose
most of their energy producing lower energy $\gamma$ and $e^\pm$. This
explains how even for very high DM masses a fair amount of photons
with energies of MeV to TeV exist.  We note that this fact makes it
possible to constrain decaying DM particles of very high masses using
Fermi-LAT observations.\footnote{We found that there is a few factor
  uncertainties in the $\gamma$-ray flux in the Fermi-LAT energy range
  using \texttt{DINT}, which was also stated in
  Ref.\,\cite{Lee:1996fp}. As it will be shown later, however, these
  uncertainties are irrelevant for the constraints on the DM
  lifetime.} Furthermore, as can be seen specially in the bottom row
of Fig.\,\ref{fig:gamma}, $\gamma$ with energies larger than
$10^{11}\,{\rm GeV}$ also survive. Consequently, the CR fluxes
observed by PAO and TA can be used to constrain such $\gamma$ fluxes.

Figure\,\ref{fig:integgamma} shows the integrated gamma flux. In this
energy range, the flux is dominated by Galactic contributions.
It is seen that the lifetime of DM is expected to be
constrained by CASA-MIA, KASCADE, and KASCADE-Grande for $m_{\rm dm}
\gtrsim 10^{9}\,{\rm GeV}$ and by TA and PAO for $m_{\rm dm} \gtrsim
10^{12}\,{\rm GeV}$.

Finally $\nu+\bar{\nu}$ fluxes are displayed in
Fig.\,\ref{fig:neutrino}. Here the Galactic contributions are shown
separately. As can be seen, the Galactic component is subdominant
compared to the extragalactic one.  As what happened in the photon
channel, neutrino fluxes in the extragalactic region are composed of
two components; prompt neutrinos from DM and secondary ones resulting
from photo-hadronic processes. We find that the secondary neutrinos
contribute much less than the prompt component. We see that the prompt
component starts to surpass observed flux or the upper bounds for DM
masses of $10^{6}\,{\rm GeV}\lesssim m_{\rm dm}\lesssim 10^{12}\,{\rm
  GeV}$. As such, this observations (upper limits) can be used to
constrain the DM lifetime in this mass range.

\subsection{Constraints on dark matter lifetime}

\begin{figure}
 \begin{center}
   \includegraphics[width=7.5cm]{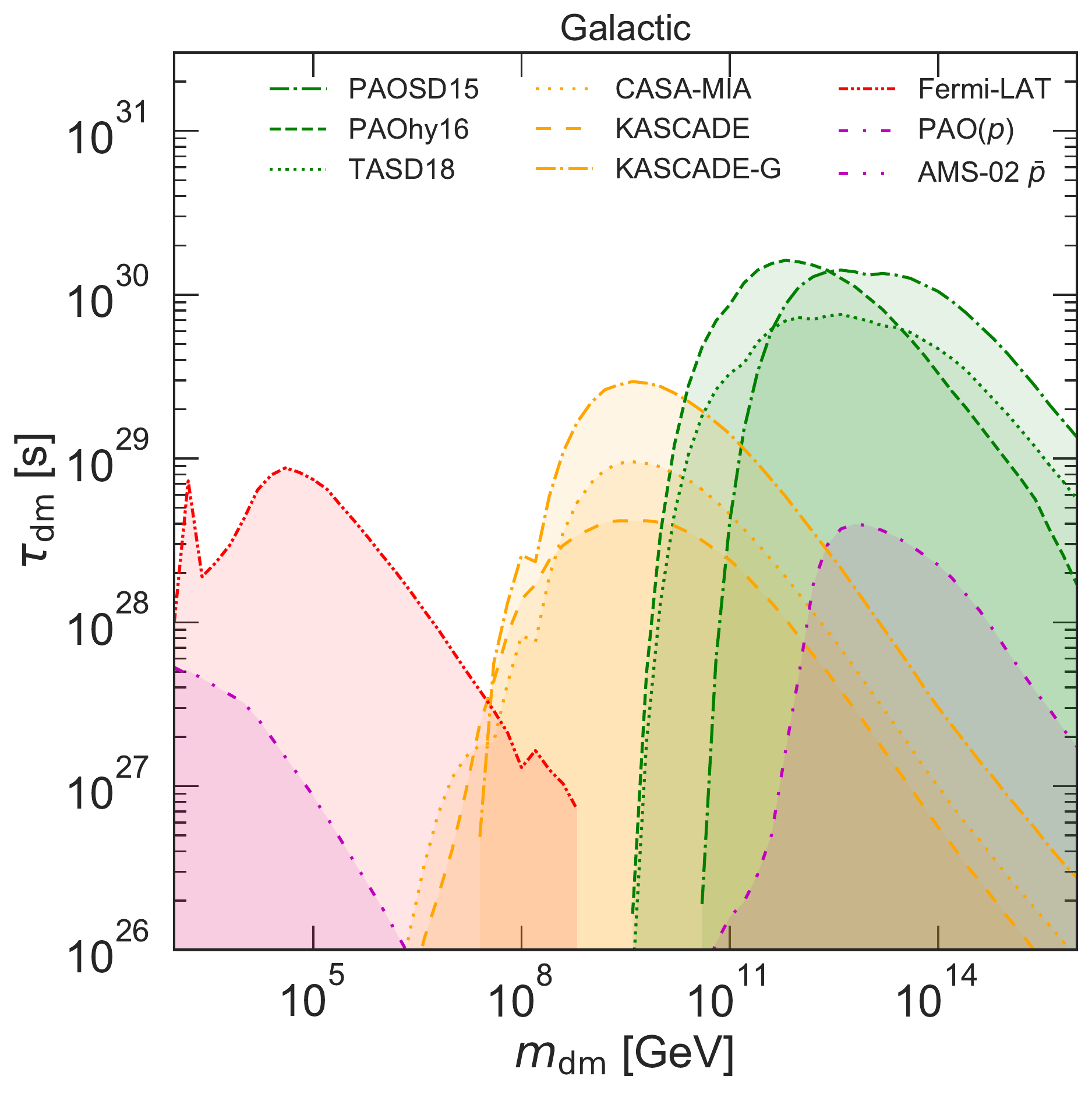}
   \includegraphics[width=7.5cm]{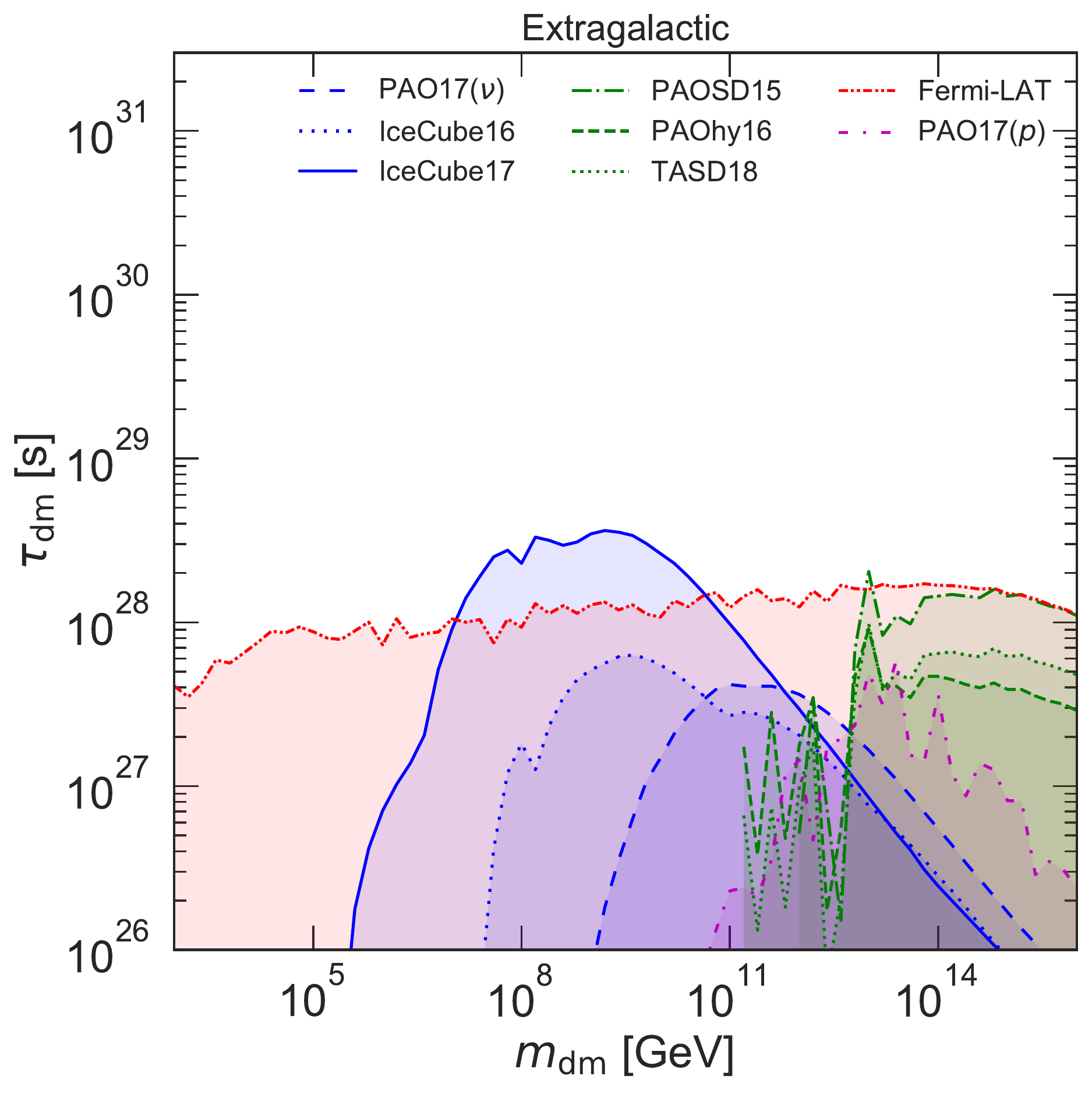}
   \caption{Conservative strong limits on the dark matter lifetime $\tau_{\rm dm}$ obtained in this work. The limits are separated according to the region in which the DM CRs were originated (left panel corresponds to the Galactic and right panel to extragalactic region). Shaded areas show     regions of the parameter space that are excluded by the CR data sets shown in the labels.}
  \label{fig:tau}
 \end{center}
\end{figure}

\begin{figure}
  \begin{center}
    \includegraphics[width=9.0cm]{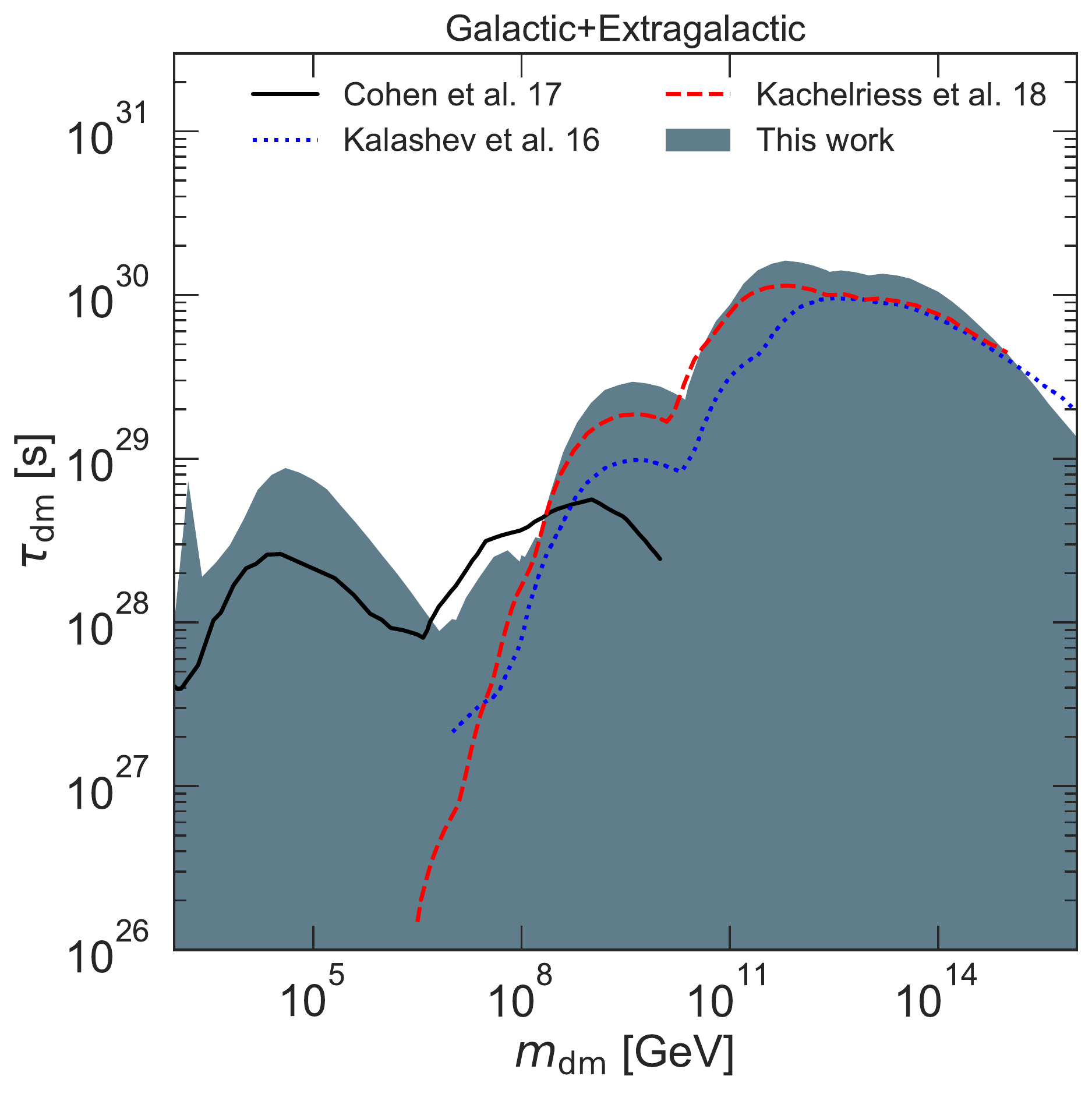}
    \caption{Same as Fig.~\ref{fig:tau}, except that here we combine the extragalactic and Galatic DM limits in the same panel. Dark blue area shows the total region of the parameter space excluded by our analysis. Limits independently obtained by recent studies~\cite{Cohen:2016uyg,Kalashev:2016cre,Kachelriess:2018rty} are also shown for comparison. }
  \label{fig:tau_comparison}
 \end{center}
\end{figure}

Using the observational data and our flux predictions, we set
conservative and robust constraints on the DM lifetime as a function
of its mass.  Figure\,\ref{fig:tau} shows the main results of our
study. To demonstrate the impact of the Galactic and extragalactic CRs
from DM, we construct lower limits on the lifetime by using
both components separately. In that figure, we derive 95\% CL limits
from Fermi-LAT and AMS-02 data while the limits from other
observations are given at the CL of each observation as shown in
Table~\ref{tab:obs}. Fig.\,\ref{fig:tau_comparison} shows a
combination of extragalactic and Galactic limits together and include
a comparison with previous results in the
literature\,\cite{Cohen:2016uyg,Kalashev:2016cre,Kachelriess:2018rty}.

The left panel of Fig.\,\ref{fig:tau} shows lower limits for the
lifetime obtained by using the Galactic fluxes, while right one is
given by the extragalactic fluxes. PAO and KASCADE-Grande give the
most stringent constraints on $\tau_{\rm dm}$ due to Galactic $\gamma$
rays. The PAO data bounds the DM lifetime $\tau_{\rm dm}\gtrsim
10^{30}\,$s for $10^{10}\,{\rm GeV}\lesssim m_{\rm dm}\lesssim
10^{15}\,{\rm GeV}$. In the mass range $10^{8}\,{\rm GeV}\lesssim
m_{\rm dm}\lesssim 10^{10}\,{\rm GeV}$, KASCADE-Grande gives the most
stringent constraints, {\it i.e.}, $\tau_{\rm dm}\gtrsim
10^{29}\,$s. We note that these results are consistent with
Ref.\,\cite{Kachelriess:2018rty}. Finally, Fermi-LAT constrains the DM
lifetime at roughly $\tau_{\rm dm}\gtrsim 10^{28}\,$s for
$10^{3}\,{\rm GeV}\lesssim m_{\rm dm}\lesssim 10^{6}\,{\rm GeV}$.
However, constrains obtained using $p+\bar{p}$ and $\bar{p}$ spectra
and PAO and AMS-02 data, respectively, are found to be weaker than
those obtained using gamma-ray observations. It is also found that the
constraints from $e^+$ flux data by AMS-02 is so weak that it is out
of the range of the plot.

The constraints obtained using extragalactic CRs are shown in the
right panel of Fig.\,\ref{fig:tau}.  It turns out that these are
weaker compared to those obtained with Galactic ones for most of the
DM mass range. The exception being the mass range of $10^{6}\,{\rm
  GeV}\lesssim m_{\rm dm}\lesssim 10^{8}\,{\rm GeV}$ where we find
that the constrains on the neutrino flux using IceCube observations
are the most stringent, {\it i.e.}, $\tau_{\rm dm}\gtrsim
10^{28}\,$s. This is consistent with limits reported in
Ref.\,\cite{Cohen:2016uyg}. It is worth noticing that Fermi-LAT gives
a constraint on the DM lifetime in the entire DM mass range.  This is
a consequence of cascading processes during the propagation CRs in the
extragalactic region. This is also in agreement with results shown in
Fig.\,3 of Ref.\,\cite{Ando:2015qda} in $m_{\rm dm}\le 10$~{TeV}
obtained through analytic modelling. This is an important consistency
check of our methods given that in this study we simulate CR particles
by using \texttt{CRPropa} instead of analytic methods described in that
reference. Reference\,\cite{Kalashev:2016cre} reports a qualitatively
similar result for $10^{7}\,{\rm GeV}\lesssim m_{\rm dm}\lesssim
10^{12}\,{\rm GeV}$, except that their bound is a factor of a few
weaker. In addition, we find that for $10^{13}\,{\rm GeV}\lesssim
m_{\rm dm}\lesssim 10^{16}\,{\rm GeV}$ the PAO constraints are
comparable to those obtained with Fermi-LAT. Although the constraints
obtained with our extragalactic predictions are found to be weaker
than those using the Galactic component, our simulations could
potentially be used in future analyses of all sky gamma-ray analyses
of, for example, tomographic cross-correlation using the local galaxy
distributions~\cite{Ando:2013xwa,Ando:2014aoa,Fornengo:2013rga,Ando:2016ang,Shirasaki:2018dkz,Hashimoto:2018ztv}.

\section{Conclusions}\label{sec:conclusions}

Using all the multi-messenger astrophysics probes --- photons,
protons, anti-protons, and neutrinos, we set constraints on the
lifetime of heavy dark matter particles in the mass ranges between
10$^4$ and 10$^{16}$~GeV.  We computed the fluxes of all the
multi-messenger probes from dark matter decays in both the Galaxy and
extragalactic halos.

The lower limits on heavy dark matter particles that we found are
summarized in Figs.~\ref{fig:tau_comparison}.  Dark
matter less massive than 10$^8$~GeV is most stringently constrained by
unresolved diffuse gamma-ray emission measured by Fermi-LAT.  For dark
matter with much heavier masses above $\sim$10$^{10}$~GeV, both gamma
rays and protons of ultrahigh energies measured with Pierre Auger
Observatory are best used to place very stringent lower limits on the
order of 10$^{30}$~s.  For masses between 10$^{8}$ and 10$^{10}$~GeV,
stringent constraints are set with KASCADE-Grande using the predicted
Galactic gamma-ray flux component.

We also found that dark matter decay yields originating in
extragalactic halos produce gamma-ray signals of GeV energies nearly
independent of dark matter mass, and hence, the Fermi-LAT diffuse
gamma-ray background are used to place constraints on the order of
10$^{28}$~s throughout the wide mass range between 10$^4$ and
10$^{16}$~GeV.  Yet, in general, the extragalactic constraints are
found to be weaker than those obtained with the Galactic component.
The only exception is the constraints obtained with the IceCube
neutrino data, which provide the best constraints on dark matter decay
in a narrow mass range around $10^7$--$10^8$~GeV.

Overall, we exclude dark matter lifetime (into $b\bar b$ final state)
of $10^{28}$~s or shorter for all the masses investigated in this
work, while the most stringent constraints reach $10^{30}$~s for very
heavy dark matter of $10^{11}$--$10^{14}$~GeV.  On the other hand,
studies on decay modes into a final state that involves leptons are
slated for a future study given that the electroweak corrections have
to be carefully assessed, which is a nontrivial problem especially for
dark matter with very heavy masses.

Although the limits derived in this work are comparable with other
existing limits in the literature, our self-consistent simulations
including extragalactic and Galactic propagation effects and all CR
species serve as an important consistency check of previous studies
and at the same time clarifies which components or modelling
assumptions have the greatest impact on the final results.

\acknowledgments

We are grateful to Daisuke Yonetoku for fruitful discussions in the early
stage of this project, Rafael Alves Batista, G\"{u}nter Sigl and
Tobias Winchen for useful discussions about the use of
\texttt{CRPropa}, Shunzo Kumano and Masanori Hirai for providing us
the codes for solving the DGLAP equations and valuable discussions,
Timothy Cohen for private communication regarding
Fig.\,\ref{fig:dNdx}, and Kohta Murase. This work was supported by JSPS KAKENHI Grant
Numbers JP17H05402, JP17K14278, JP17H02875, JP18H05542 and Sakigake 2018 Project of Kanazawa University (KI). SA and OM were supported by World Premier International Research Center Initiative (WPI Initiative), MEXT, Japan and by JSPS KAKENHI Grant
Numbers JP17H04836, JP18H04340 and JP18H04578. MA was JSPS KAKENHI Grant Numbers JP17H06362 and 
the JSPS Leading Initiative for Excellent Young Researchers program.

\end{document}